\documentclass[12pt,letterpaper]{article}
\RequirePackage{ifpdf}
\pdfoutput=1
\usepackage{jheppub}
\usepackage{epsfig}
\usepackage[latin1]{inputenc}
\usepackage{bbm,amsfonts}
\usepackage{graphicx}
\usepackage{amssymb,amsmath,amsfonts,mathtools}
\usepackage{fancybox}
\usepackage{enumerate}
\usepackage{dsfont}
\usepackage{verbatim}
\usepackage{wrapfig}
\usepackage{slashed}

\author[a]{Marco S. Bianchi}
\author[b]{and Andrea Mauri}
 
\affiliation[a]{Center for Research in String Theory - School of Physics and Astronomy Queen Mary University of London, Mile End Road, London E1 4NS, UK}
\affiliation[b]{Dipartimento di Fisica, Universit\`a degli Studi di Milano-Bicocca and INFN, Sezione di Milano-Bicocca, Piazza della Scienza 3, I-20126 Milano, Italy}

\emailAdd{m.s.bianchi@qmul.ac.uk}   
\emailAdd{andrea.mauri@mi.infn.it} 
\preprint{QMUL-PH-17-17}
  
\abstract{In ABJ(M) theory a generalized cusp can be constructed out of the 1/6-BPS Wilson line by introducing an angle $\varphi$ in the spacial contour and/or an angle $\theta$ in the internal R-symmetry space. The small angles limits of its anomalous dimension are controlled by corresponding Bremsstrahlung functions.
In this note we compute the internal space $\theta$-Bremsstrahlung function to four loops at weak coupling in the planar limit. 
Based on this result, we propose an  all order conjecture for the $\theta$-Bremsstrahlung function.
}

\title{ABJM $\theta$-Bremsstrahlung at four loops and beyond}


\newcommand{\be}{\begin{equation}}
\newcommand{\ee}{\end{equation}}
\newcommand{\beq}{\begin{equation}}
\newcommand{\eeq}{\end{equation}}
\newcommand{\bea}{\begin{eqnarray}}
\newcommand{\eea}{\end{eqnarray}}
\newcommand{\ena}{\end{eqnarray}}
\newcommand {\non}{\nonumber}

\renewcommand{\a}{\alpha}
\renewcommand{\b}{\beta}

\renewcommand{\d}{\delta}
\renewcommand{\th}{\theta}

\newcommand{\pa}{\partial}
\newcommand{\g}{\gamma}
\newcommand{\G}{\Gamma}

\newcommand{\e}{\epsilon}

\renewcommand{\l}{\lambda}

\newcommand{\m}{\mu}

\newcommand{\n}{\nu}

\newcommand{\p}{\pi}

\newcommand{\s}{\sigma}

\renewcommand{\t}{\tau}

\def\Tr{\textrm{Tr}}

\numberwithin{equation}{section}

\def\clock{{\count0=\time
           \divide\count0 60
           \ifnum\count0<10 0\fi\the\count0
           \multiply\count0 -60 \advance\count0 \time
           :\ifnum\count0<10 0\fi \the\count0
         }}
\newcommand{\timestamp}{{\small\vbox{\hbox{\tt\jobname.tex}
\hbox{\the\day/\the\month/\the\year, \clock}}}}
\begin{document}

\maketitle
\allowdisplaybreaks

\section{Introduction}

ABJM theory in three dimensions \cite{Aharony:2008ug,Aharony:2008gk} is likely to be solvable, at least in the planar limit, as it is believed to be the case for its four-dimensional cousin, maximally supersymmetric Yang-Mills theory.
Several exact results for certain observables in the ABJM model are already available from established techniques such as integrability \cite{MZ,Gaiotto:2008cg,GV,BR,BGR} and localization \cite{Kapustin:2009kz,Marino:2009jd,Drukker:2010nc}, paralleling progress in four-dimensional ${\cal N}=4$ SYM.
In particular, integrability allows in principle to solve exactly for the anomalous dimensions of composite operators, in the planar limit \cite{Beisert:2010jr}, by mapping the dilatation operator of the theory to the Hamiltonian of an integrable spin chain.
Furthermore certain supersymmetric theories (for instance ${\cal N}=4$ SYM in four dimensions and ${\cal N}=2$ Chern-Simons-matter theories in three dimensions, including ABJM) can be defined on curved compact manifolds where the path integral of the theory localizes onto a matrix model. Certain supersymmetric observables of these theories, notably circular Wilson loops, can be computed as matrix model averages and if the resulting matrix model can be solved, this provides exact formulae for their expectation values \cite{Pestun:2007rz}.

These techniques seem to apply to different sectors of the given theory. However, in ${\cal N}=4$ SYM a particular object was found that lies in both the ranges of applicability of localization and integrability \cite{Correa:2012at}.
This is the so-called Bremsstrahlung function.
Such an object governs the small angle $\varphi$ expansion of the cusp anomalous dimension $\G_{cusp}(\varphi)$ that, in turn, controls the short distance divergences of a Wilson loop near a cusp, according to the universal behaviour $\langle WL \rangle \sim \exp{(-\G_{cusp} \log{\frac{\Lambda}{\mu}})}$ (with $\Lambda$ and $\mu$  IR and UV cutoffs, respectively).
In formulae the Bremsstrahlung function is defined as
\begin{equation}
\G_{cusp}(\varphi) = -\varphi^2\, B + {\cal O}(\varphi^4)
\end{equation}
In a conformal field theory, this function can be shown to also govern the energy radiated by a massive probe (a quark), moving at a velocity $v$, undergoing a deviation in its trajectory by an angle $\varphi$, in the small angle limit \cite{Correa:2012at}
\begin{equation}
\Delta E \sim B\, \int dt\, |\dot v(t)|^2
\end{equation}
hence the name Bremsstrahlung function.

More precisely, we consider supersymmetric extensions of ordinary Wilson loops, given as the holonomy of generalized connections that include also couplings to matter. Consequently we can consider a cusped Wilson loop which depends on two parameters, $\varphi$ representing the geometric angle between the two Wilson lines meeting at the cusp, and an internal space angle $\theta$ describing the change in the orientation of the couplings to matter between the two rays \cite{Drukker:1999zq,Drukker:2011za}, thus defining a generalized cusp $\G_{cusp}(\varphi,\theta)$. Hence one derives the small angles expansion of $\G_{cusp}$ 
\begin{equation}\label{eq:BPS}
\Gamma_{cusp} (\theta, \varphi) \sim (B^{\theta}\,\theta^2 - B^{\varphi}\,\varphi^2)   
\end{equation}
where $B^{\theta}$ and $B^{\varphi}$ are two a priori distinct Bremsstrahlung functions, associated to the respective angles. They are both expressed as a functions of the coupling constant of the theory, e.g.~the 't Hooft coupling $\lambda$, in the planar limit.
In certain remarkable cases, the generalized cusp satisfies a BPS condition (this happens for $\theta^2=\varphi^2$ in all known examples), where some amount of supersymmetry is conserved. As a consequence the cusp anomalous dimension vanishes in such a situation, which in turn forces the Bremsstrahlung functions to coincide, since the BPS condition has to hold in the small angle limit as well.

We stress that in principle the Bremsstrahlung function is not a supersymmetric quantity and hence cannot be localized. Nonetheless, in the context of ${\cal N} = 4$ SYM a prescription was devised in the seminal paper \cite{Correa:2012at}, so as to extract an exact formula for this non-BPS observable in terms of BPS loops which can be determined explicitly via localization. 
Remarkably, the very same result can be obtained with an integrability based approach. This deals with an exact set of TBA equations \cite{Bombardelli:2009ns,Gromov:2009bc,Arutyunov:2009ur} describing the generalized cusp \cite{Correa:2012hh,Drukker:2012de} in the near-BPS limit \cite{Gromov:2012eu}. This is done by considering the spectral problem for certain operators inserted at the tip of the cusp, which is mapped to an integrable spin chain with reflecting boundary conditions. Moreover, the use of the quantum spectral curve techniques \cite{Gromov:2013pga,Gromov:2014caa} has allowed to obtain results away from the BPS point and in a number of generalized settings \cite{Gromov:2012eu,Gromov:2013qga,Gromov:2015dfa}. 

Since the generalized cusps constructed with supersymmetric Wilson loops (and their small angle limits) have proven to be such a fruitful playground in the search for exact results in ${\cal N}=4$ SYM, in this note we aim at its extension to three-dimensional ABJM theory.
In this setting, a first stark difference emerges, with respect to the four-dimensional case. 
In ${\cal N}=6$ Chern-Simons-matter theories one can consider two structurally different supersymmetric Wilson loops: the 1/6-BPS \cite{Berenstein:2008dc,Drukker:2008zx,Chen:2008bp,Rey:2008bh} and the 1/2-BPS \cite{Drukker:2009hy}, by supplementing the gauge connection with some coupling to the matter fields of the theory. 
In particular, the first are bosonic objects, in the sense that they are constructed as the holonomy of a connection containing the gauge field (as in the ordinary case) and a coupling to a bi-scalar operator.
On the contrary, the latter also feature a coupling to fermion fields which can be elegantly embedded in generalized superconnections (in the sense that they are super-Lie algebra valued) whose holonomy gives rise to 1/2-BPS loop operators. 
In particular, these operators are holographically dual to fundamental strings in $AdS_4\times CP^3$. Moreover, we recall that he 1/2 BPS Wilson loop is cohomologically equivalent to a linear combination of 1/6 BPS ones, meaning that their difference is annihilated by a supercharge. For circular Wilson loops this property translates to a relation between the respective expectation values computed via localization. 

Still, the fact that they preserve different amounts of supersymmetry allows for the construction of different non-BPS observables (i.e.~generalized cusps) from them. 
Indeed, generalized cusps formed with 1/6-BPS rays or 1/2-BPS rays are actually different \cite{Griguolo:2012iq,Lewkowycz:2013laa} and, consequently, different Bremsstrahlung functions can be defined and potentially evaluated exactly. 
The construction of such cusps in ABJM is summarized as follows.
Taking a pair of straight lines meeting at an angle $\varphi$, a cusp is introduced, which can be generalized \cite{Drukker:1999zq,Drukker:2011za} via an additional deviation in the R-symmetry space of couplings to the matter fields, by an internal angle $\theta$ \cite{Griguolo:2012iq}. This configuration breaks supersymmetry in general and consequently the expectation value of the Wilson loop is divergent and the operator acquires a cusp anomalous dimension. This is in general a function of the two angles (and of the coupling and gauge group ranks of the theory). As recalled in \eqref{eq:BPS}, the first coefficients in the small angles Taylor expansion of the cusp anomalous dimension are called Bremsstrahlung functions and are in principle two unrelated objects for the two angles.

For the cusp constructed with 1/2-BPS rays a BPS condition is satisfied for $\varphi=\theta$, where some supersymmetry is preserved and the divergence cancels. As a result, the coefficients of the small angle expansion for $\varphi$ and $\theta$ are opposite and one can define a unique Bremsstrahlung function $B_{1/2}$.
At a difference, no BPS condition seems to hold for the cusp built out of two 1/6-BPS lines. Therefore two different Bremsstrahlung functions are present in this case $B_{1/6}^{\varphi}$ and $B_{1/6}^{\theta}$.

Paralleling the success in ${\cal N}=4$ SYM, the Bremsstrahlung functions of ABJM theory are amenable of exact computations.
This project has been partially (and with some degree of conjecture) attained, and some proposals exist for them, relating their expression to the expectation value of 1/6-BPS Wilson loops $\langle W_{n} \rangle$ wound $n$ times around the great circle, which can be computed exactly thanks to localization.
The precise state of the art for these Bremsstrahlung functions is summarized as follows.
\begin{itemize}
\item For $B_{1/2}$ a conjecture was put forward \cite{Lewkowycz:2013laa,Bianchi:2014laa} on its exact expression in terms of 1/6-BPS Wilson loops which are computable exactly via localization (see also \cite{Bianchi:2017ozk}). It agrees with explicit computations at weak \cite{Griguolo:2012iq,Bianchi:2014laa,Bianchi:2017svd}, up to three loop order, and strong coupling \cite{Forini:2012bb,Correa:2014aga,Aguilera-Damia:2014bqa}, up to the subleading order.
The proposed formula reads
\begin{equation}
B_{1/2} = -\frac{i}{8\pi}\, \frac{\langle W_{1}\rangle-\langle \hat W_{1}\rangle}{\langle W_{1}\rangle+\langle \hat W_{1}\rangle}
\end{equation}
and is valid in the ABJM limit where the gauge group ranks are equal.
\item A proposal for the exact $B_{1/6}^{\varphi}$ appeared in \cite{Lewkowycz:2013laa}, which passes a strong coupling check \cite{Correa:2014aga,Aguilera-Damia:2014bqa} up to the subleading order and a weak coupling two-loop computation \cite{Griguolo:2012iq,Bianchi:2014laa}.
This proposal reads
\begin{equation}\label{eq:Bphi}
B_{1/6}^{\varphi} = \frac{1}{4\pi^2}\, \partial_n\, |W_n|\, \bigg|_{n=1}
\end{equation}
and was again derived in the equal ranks limit.
\item The expression for $B_{1/6}^{\theta}$ was related to a certain supersymmetric Wilson loop expectation value in \cite{Bianchi:2014laa,Correa:2014aga}.
This is a circular Wilson loop preserving 2 supersymmetries of the theory which is evaluated on a latitude contour on $S^2$ and with a nontrivial profile for the coupling to the scalars in the connection.
Unfortunately, the latter is not known exactly, thus preventing from deriving an all order expression. We remark that a few perturbative orders were computed at weak coupling (up to two-loop order \cite{Bianchi:2014laa}), but its computation at strong coupling, where ABJM theory is dual to type IIA string theory on $AdS_4\times CP^3$, is lacking and it remains unclear how to approach it \cite{Aguilera-Damia:2014bqa}.
Hence, no explicit exact result is available for this quantity yet.

\item Finally, no integrability computation exists for any of these functions, thus far, though progress has been made in this direction \cite{Cavaglia:2014exa,Bombardelli:2017vhk}. This constitutes a stark difference with respect to ${\cal N}=4$ SYM in four dimensions. Such a result would be considerably desirable, because it would potentially relate an integrability based computation to another exact formula derived by other means (localization in this case). This would grant a firmer handle on the interpolating $h(\lambda)$ function which appears in all integrability computations in ABJM and whose exact value is thus far only conjectured \cite{Gromov:2014caa}.
\end{itemize}

As clear from the summary above, the Bremsstrahlung function associated to the internal angle $B_{1/6}^{\theta}$ is the least understood object in this picture.
In this note we aim at filling this gap and focus on this quantity. We start by reviewing the basics of the construction of 1/6-BPS Wilson lines in ABJM and their generalized cusp configuration in section \ref{sec:cusp}.
Then the logic behind our analysis of $B_{1/6}^{\theta}$ is as follows. 

We start trying to get some mileage by computing this quantity in the weak coupling approximation. 
The two-loop result was already computed in \cite{Griguolo:2012iq,Bianchi:2014laa} and since only even loops provide divergent contributions to this object we address the computation at the next relevant order which is four loops.
Due to this high perturbative order the computation is rather involved, nevertheless it turns out to be completely within the reach of modern technologies. The details of such a calculation are collected in the appendices in order not to shadow the main results of the paper with technicalities. The main strategy underpinning the calculation is spelled out in section \ref{sec:computation}.
We restrict our approach to the planar limit, but we allow for generic gauge group ranks (i.e.~we keep them distinct as in the ABJ generalization \cite{Aharony:2008gk}).

The result of this computation is reported in formula \eqref{eq:bremsstrahlung} and constitutes one of the main achievements of the paper. This is the starting point of our subsequent argumentations which are developed in section \ref{sec:conj} and briefly outline here.
The perturbative result for $B_{1/6}^{\theta}$ and its comparison with $B_{1/6}^{\varphi}$ suggest a simple relation between the two observables
\begin{equation}\label{eq:main}
B_{1/6}^{\varphi} = 2\, B_{1/6}^{\theta}
\end{equation}
We remark that we have verified such a relation up to fourth order at weak coupling. The agreement is non-trivial since it occurs for all the coefficients of the different powers of the gauge group ranks $N_1$ and $N_2$ that we keep distinct.
This provides quite a compelling hint at the validity of \eqref{eq:main} beyond four loops and we conjecture that is holds to all orders.
If true, relation \eqref{eq:main} entails that $B_{1/6}^{\theta}$ can in turn be computed exactly and that it is related to the expectation value of a multiply wound 1/6-BPS circular Wilson loop, as is $B_{1/6}^{\varphi}$, as recalled in \eqref{eq:Bphi}.
Hence, this study completes the picture of the exact computation of the Bremsstrahlung functions for ABJM theory, in the planar approximation, by supplying a conjecture for the last missing ingredient: $B_{1/6}^{\theta}$.
In particular, this allows its straightforward computation at strong coupling, where as of today, no string computation is available for this object yet \cite{Correa:2014aga,Aguilera-Damia:2014bqa}.

\section{The cusp}\label{sec:cusp}

\subsection{The 1/6-BPS generalized cusp in ABJM}

We start by defining the generalized cusped Wilson loop constructed with 1/6-BPS rays.
We consider the ABJM model with gauge groups $U(N_1)_k\times U(N_2)_{-k}$ with Chern-Simons level $k$. 
The Lagrangian of the theory as well as its Feynman rules which we use for perturbative computations are collected in appendices \ref{app:ABJM} and \ref{app:FeynmanRules}.
We restrict to the planar limit $N_1,\, N_2 \gg 1$ in the weak coupling regime $k \gg N_1,\, N_2$.
The gauge connections for the gauge groups $A,\, \hat{A}$, along with complex scalars $C_I, \bar C^J$ and fermions $\psi^I, \bar \psi_J$ (where $I,J = 1,\dots 4$) transforming in the bi-fundamental representation, constitute the field content of the theory. 
The 1/6-BPS Wilson loop \cite{Drukker:2008zx,Chen:2008bp,Rey:2008bh} in Euclidean space reads
\begin{equation}\label{eq:WL}
W_{1/6}[\Gamma] = \frac{1}{N_1}\, \Tr \left[ \,\textrm{P}\exp{ -i \int_\Gamma d\tau~ \left( A_{\mu} \dot x^{\mu}-\frac{2 \pi i}{k} |\dot x|\, M_{J}^{\ \ I} C_{I}\bar C^{J} \right)(\tau) } \right] 
\end{equation}
on a contour $\Gamma$ describing a cusp at an angle $\varphi$
\begin{equation}\label{eq:contour}
\Gamma: \ \ \ \ x^{0}=0 \ \ \ \ \ x^{1}=s\cos\frac{\varphi}{2} \ \ \ \ \  x^{2}=|s|\sin\frac{\varphi}{2}\ \ \ \ \  -\infty\le s\le \infty
\end{equation}
We can introduce an additional angle $\theta$ in the internal space, by taking different coupling matrices $M$ with the scalars on the two edges of the cusp
\begin{equation} \label{eq:Mmatrices}
M_{1J}^{\ \ I}=\mbox{\small $\left(
\begin{array}{cccc}
 -\cos \frac{\theta }{2}& -\sin \frac{\theta }{2} & 0 & 0 \\
 -\sin \frac{\theta }{2}& \cos\frac{\theta }{2} & 0 & 0 \\
 0 & 0 & -1 & 0 \\
 0 & 0 & 0 & 1
\end{array}
\right)$}
\ \ \ \ \mathrm{and}\ \ \ \  M_{2J}^{\ \ I}=\mbox{\small $\left(
\begin{array}{cccc}
 -\cos \frac{\theta }{2} & \sin \frac{\theta }{2} & 0 & 0 \\
 \sin \frac{\theta }{2} & \cos\frac{\theta }{2} & 0 & 0 \\
 0 & 0 & -1 & 0 \\
 0 & 0 & 0 & 1
\end{array}
\right)$}
\end{equation}
This configuration breaks the supersymmetries of the individual straight line Wilson loops, which is not restored even for special values of the angles, but $\varphi=\theta=0$.
As a consequence the vacuum expectation value of the Wilson loop develops ultraviolet divergences and the operator has an anomalous dimension $\Gamma_{1/6}(\varphi,\theta)$ which is in general a function of the angles, according to the universal behaviour
\begin{equation}
\label{espo}
\langle W_{cusp} \rangle = e^{-\Gamma_{cusp}(k,N,\varphi,\theta)\, \log \frac{\Lambda}{\mu}} + finite
\end{equation}
where $\Lambda$ is an IR cutoff and $\mu$ stems for the renormalization scale.
The divergences associated to the cusp singularity of Wilson loops exponentiate thanks to general exponentiation theorems for non-local operators \eqref{espo} \cite{Korchemsky:1987wg}. More precisely, their expectation value  can be written as the  exponential of the  sum of all two--particle irreducible diagrams \cite{Dotsenko:1979wb,Gatheral:1983cz}. Thus, we expect the 1/6-BPS Wilson line in ABJM to respect the usual exponentiation pattern, allowing us to define and compute the anomalous dimension for the cusp, according to standard text-book procedures. Our four-loop result, that we derive in the following, explicitly confirms the correctness of this picture (see further discussion in section \ref{sec:result}).

\subsection{The 1/6-BPS Bremsstrahlung functions}

In the limit where the generalized cusp angles $\varphi$ and $\theta$ are small, the cusp anomalous dimension $\Gamma(\theta,\varphi)$ can be Taylor expanded in even powers of them. The coefficients of the lowest orders in $\varphi$ and $\theta$ define the Bremsstrahlung functions \cite{Correa:2012at}.

The generalized cusp, associated to the 1/6-BPS Wilson lines of the ABJM theory, does not satisfy a BPS condition at $\varphi=\pm \theta$ (as for instance the cusp constructed with 1/2-BPS rays does) and consequently the small angle expansion reads
\begin{equation}\label{eq:Brems}
\Gamma_{1/6}(k,N_1,N_2,\varphi,\theta) = B_{1/6}^{\theta}(k,N_1,N_2)\, \theta^2 - B_{1/6}^{\varphi}(k,N_1,N_2)\, \varphi^2 + \dots
\end{equation}
where $B_{1/6}^{\theta}$ and $B_{1/6}^{\varphi}$ are a priori two different functions of the coupling constant $k^{-1}$ and the number of colors $N_1$ and $N_2$ only.

Remarkably, a proposal was derived for an exact expression for $B_{1/6}^{\varphi}$ in \cite{Lewkowycz:2013laa}. This was originally obtained in the planar limit and in the ABJM limit of equal gauge group ranks. The argument relates this to the one-point function of the stress-energy tensor, 
which in turn was argued to be connected to the entanglement entropy on a spherical region enclosing a Wilson line insertion.
Finally, the computation of this entanglement entropy was expressed after some steps in terms of the expectation value of a multiply wound 1/6-BPS Wilson loop $\langle W_n \rangle$, which can be computed exactly via localization \cite{Klemm:2012ii}.
In formulae
\begin{equation}
B_{1/6}^{\varphi} = \frac{1}{4\pi^2}\, \partial_n |\langle W_n \rangle |\, \bigg|_{n=1}
\end{equation}
whose weak and strong coupling expansions read
\begin{align}
B_{1/6}^{\varphi} &= \frac{\lambda^2}{2} - \frac{\pi^2\,\lambda^4}{2} + \frac{47\pi^4\,\lambda^6}{72} - \frac{17\pi^6\, \lambda^8}{18} + {\cal O}\left(\lambda^{10}\right) \qquad\qquad \lambda\equiv \frac{N}{k}\ll 1\\
B_{1/6}^{\varphi} &= \frac{\sqrt{\lambda}}{2\sqrt{2}\pi} - \frac{1}{4\pi^2} + \left( \frac{1}{4\pi^3} + \frac{5}{96\pi} \right) \frac{1}{\sqrt{2\lambda}} + {\cal O}\left(\lambda^{-3/2}\right) \qquad \lambda\equiv \frac{N}{k}\gg 1
\end{align}
where $\lambda\equiv\frac{N}{k}$ is the 't Hooft coupling of ABJM.
On the other hand, the Bremsstrahlung function $B_{1/6}^{\theta}$ was related to the expectation value of a supersymmetric circular Wilson loop evaluated on a latitude contour in the $S^2$ sphere, that is displaced by an angle from the maximal circle, and potentially by an additional internal angle $\alpha$ in the R-symmetry space. We refer the readers to \cite{Cardinali:2012ru}, for the full details of its construction. Remarkably, the expectation value of such a Wilson loop appears to only depend on a certain combination of these parameters, $\nu= \sin{2\a} \cos{\th_0}$ \cite{Bianchi:2014laa}, where the un-deformed Wilson loop recovered at $\nu=1$ corresponds to the 1/6-BPS circular Wilson loop.
The relation between $B_{1/6}^{\theta}$ and this object was hinted at in \cite{Bianchi:2014laa} and  developed more formally in \cite{Correa:2014aga}, paralleling an analogous derivation for ${\cal N}=4$ SYM \cite{Correa:2012at}.
Such a relation eventually states that the $B_{1/6}^{\theta}$ Bremsstrahlung function is obtained as the derivative of the latitude Wilson loop expectation value $\langle W(\nu)\rangle $ with respect to the deformation parameter $\nu$
\begin{equation}
B_{1/6}^{\theta} = \frac{1}{4\pi^2}\, \partial_\nu\,\log\, \left| \langle W(\nu)\rangle \right|  ~ \Big|_{\nu=1}
\end{equation}
Such a formula was verified to hold at first order (two loops) at weak coupling in \cite{Bianchi:2014laa}.
Unfortunately, unlike the ${\cal N}=4$ SYM case, the latitude Wilson loop has not been given an exact expression via localization yet, thus impeding the derivation of an exact formula for the $\theta$-Bremsstrahlung function.

\section{Strategy of the computation}\label{sec:computation}

In this section we sketch the computation of the $\theta$-Bremsstrahlung function up to order 4 at weak coupling. 
The computation consists in few successive steps, which we summarize here and describe in more details in a number of dedicated appendices. It proceeds as a standard perturbation theory Feynman diagram expansion and evaluation:
\begin{itemize}
\item First, we select only the diagrams that actually contribute to $B_{1/6}^{\theta}$, among the full set of possible Feynman graphs.
\item The second step consists in writing down algebraic expressions for the diagrams  using the Feynman rules of the theory of appendix \ref{app:FeynmanRules}. In this step the diagrams are Fourier transformed to momentum space.
\item In the third step, the momentum space integrals of the diagrams are manipulated using automatized implementations of integration by parts (IBP) techniques.  The outcome of this step is that each diagram is mapped to a linear combination on a basis of master integrals.
\item In the fourth and final step, the master integrals are evaluated in $d=3-2\e$ dimensions, leading to a final result for each diagram in the form of an $\e$-expansion.  From the expansion of the total result, as explained below,  we can finally read $B_{1/6}^{\theta}$.
\end{itemize}
We now analyze each step in more details, starting from the selection of the diagrams. 

\subsection{The diagrams}\label{sec:ovdiag}

To compute  $B_{1/6}^{\theta}$ we do not need the full set of diagrams generated by the perturbative expansion of $\langle W_{1/6}[\Gamma]\rangle$, since several simplifications take place. First, it can be argued that only even perturbative orders are non-vanishing, on the basis of the Feynman rules of the ABJM theory only \cite{Rey:2008bh}. Hence we consider two and four-loop diagrams only.

At four loops,  as argued in \cite{Bianchi:2017svd}, for the $\theta$-Bremsstrahlung function it suffices to focus on the contributions which contain $\theta$, while setting the geometric angle to 0. The first restriction entails the following simplifications. Terms containing the internal angle $\theta$ can only come from the couplings to the scalar fields \eqref{eq:Mmatrices}. Hence we can consider only the subset of Feynman diagrams with insertions of scalar fields.
In particular, these terms arise (up to 4 loops) from the following traces 
\begin{equation}\label{eq:traces}
\Tr (M_1 M_2) = \Tr (M_1^3 M_2) = \Tr (M_1 M_2^3) = 4\cos^2\frac{\theta}{2} \equiv 4 C^2_\theta
\end{equation}
As the angle $\theta$ appears in the following computation mainly in the form above, we have defined the shorthand notation $C_\theta$ which will appear ubiquitously in the results.
This means that at least two bi-scalar insertions have to lie on different edges of the cusp, contributing with a $M_1$ and $M_2$ factors in the traces.
 
Moreover, we anticipate that according to the general prescription by Korchemsky and Radyushkin \cite{Korchemsky:1987wg} we can restrict to the evaluation of the 1PI contribution $V(\theta,\varphi)$, which is given by connected corrections that do not entirely lie on either side of the cusp, but rather stretch between the two cusp legs. We review this argument in more detail below. The knowledge of $V(\theta,\varphi)$ is sufficient to completely reconstruct the full gauge invariant cusp $W(\theta,\varphi)$ expectation value (which already includes the subtraction of terms arising from the contour being open \cite{Griguolo:2012iq,Bianchi:2017svd}).  In fact, according to the prescription of   \cite{Korchemsky:1987wg}, the full result can  be obtained by subtracting the 1PI total at vanishing angles
\begin{equation}
\label{eq:KorchRad}
\log W(\theta,\varphi) =\log V(\theta,\varphi) - \log V(0,0)
\end{equation}
We stress that the additional contribution from $V(0,0)$ is by definition independent of $\theta$, hence it does not contribute to the $\theta$-Bremsstrahlung function and can consequently be disregarded from the beginning.

\begin{figure}[h]
\centering
\includegraphics[scale=0.3]{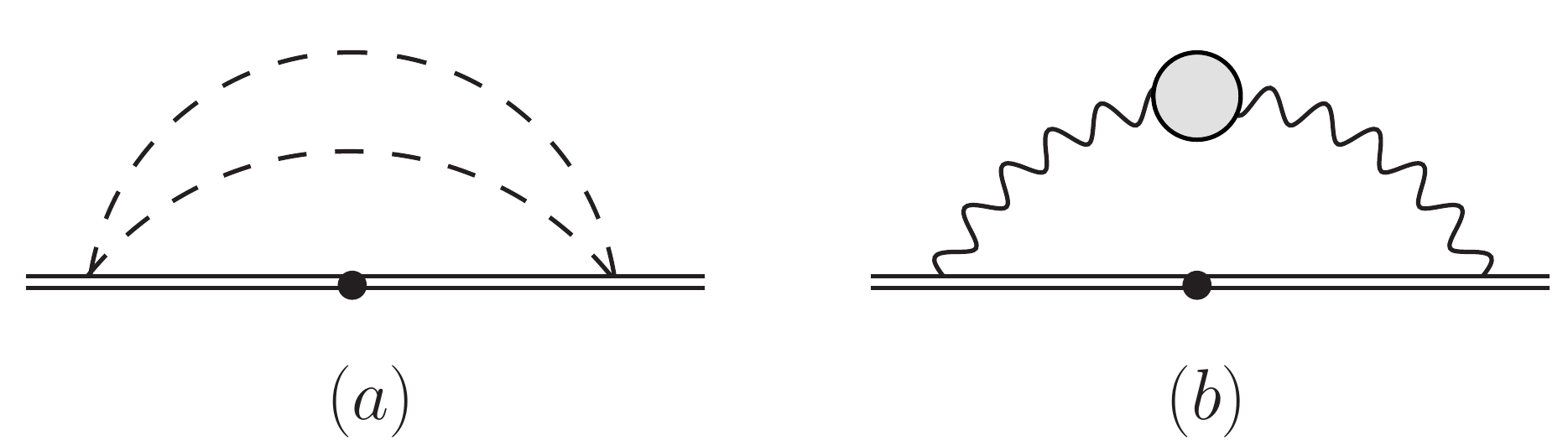}
\caption{List of two-loop diagrams contributing to the four-loop $\theta$-Bremsstrahlung function.} \label{fig:2loops}
\end{figure}

Summarizing,  in order to compute the $\theta$-Bremsstrahlung function, we have to consider the 1PI two- and four-loop diagrams at $\varphi=0$ with   $\theta$ dependent factors. 
The former contain just one contribution with $\theta$ dependence sketched in Figure \ref{fig:2loops}(a), but we also need the additional $\theta$ independent diagram arising from the gluon 1-loop self-energy \ref{fig:2loops}(b), in order to consistently extract the perturbative logarithm at four loops. The double line here stands for the Wilson line while solid, curly and dashed lines represent respectively fermion, vector and scalar fields. 

\begin{figure}[h]
\centering
\includegraphics[scale=0.3]{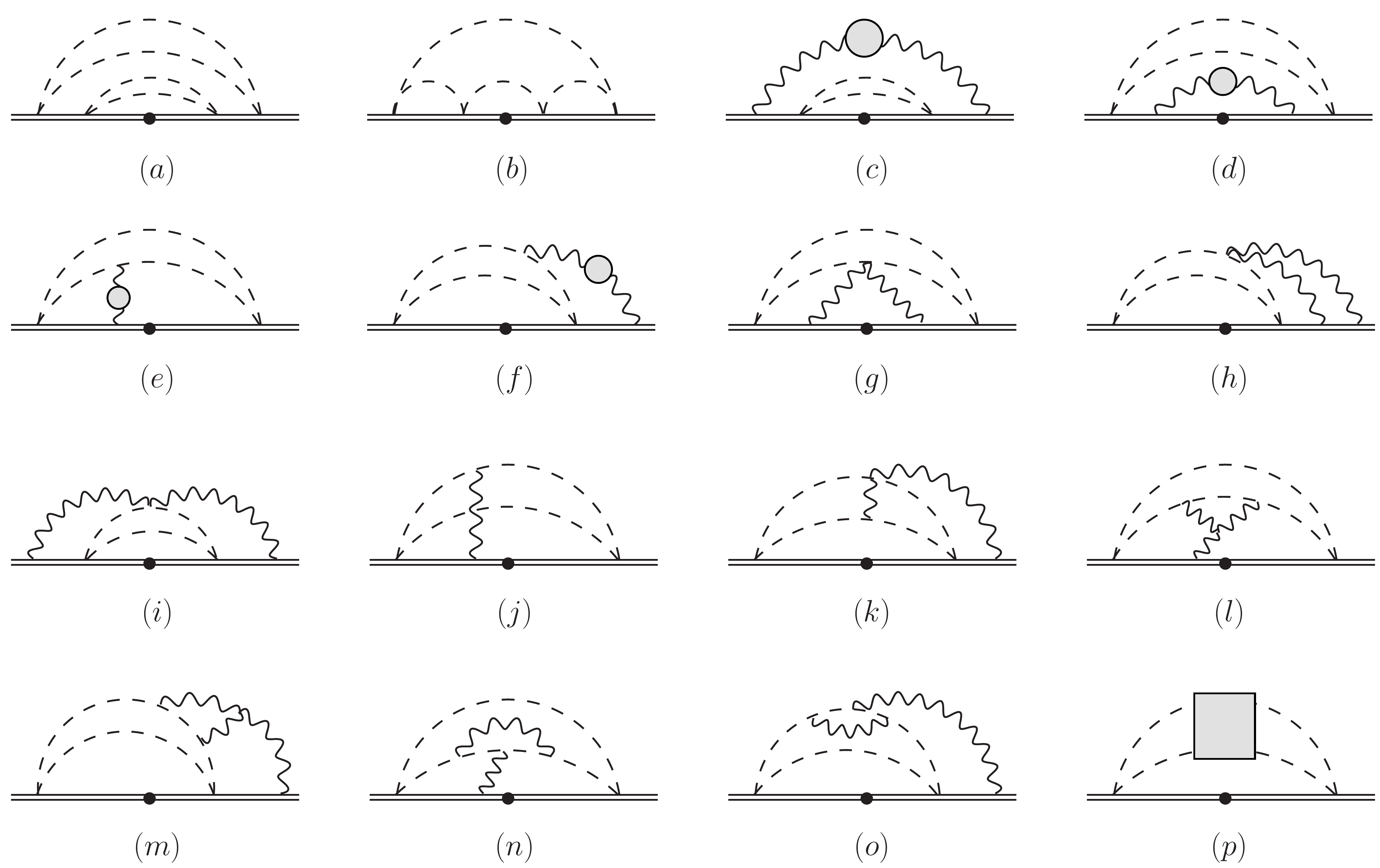}
\caption{List of four-loop diagrams contributing to the four-loop $\theta$-Bremsstrahlung function.} \label{fig:4loops}
\end{figure}

\noindent 
The four-loop diagrams are depicted in Figure \ref{fig:4loops}.
\noindent 
Some comments are in order.
First, we stress that some additional diagrams are not shown in the Figure, since they can be seen to vanish identically.
Indeed, we recall that in Chern-Simons-matter theories remarkable simplifications arise already when the Wilson loop contour lies in a two-dimensional plane, let alone when the contour is a one-dimensional line, as in our case thanks to the choice $\varphi=0$. In fact, this forces the vanishing of various tensor contractions, by virtue of the ubiquitous antisymmetric Levi-Civita tensors arising in Chern-Simons theory, for instance from gauge propagators \eqref{treevector} and cubic vertex (\ref{gaugecubic}).
Specifically, we remark that whenever an odd number of antisymmetric tensors $\varepsilon_{\mu \nu \rho}$ emerges from the algebra of a diagram, then it identically vanishes. This vanishing occurs as one can always reduce a product of an odd number of Levi--Civita tensors to a single one whose indices have to be contracted with three external vectors. The latter all come from the WL contour and hence, lying on a plane, are not linearly independent and always give vanishing expressions when contracted with the $\varepsilon$ tensor. This argument is even stronger on a one-dimensional contour where all external vectors are proportional and hence it suffices  that a pair of them are contracted with an antisymmetric tensors to obtain a vanishing result.
This observation drastically reduces the number of contributions to be considered (and in particular can be shown to force all odd loop orders to vanish, using the Feynman rules of the theory and the definition of the 1/6-BPS Wilson loop).

Concerning the diagrams in Figure \ref{fig:4loops}, the dot on the Wilson line represents one of the possible positions of the cusp point. Indeed, for some of the diagrams the cusp point can be chosen to stay in different inequivalent sites on the Wilson line. We consider here only the configurations where a $\theta$ dependent factor \eqref{eq:traces}  is generated and sum over them.

Diagrams (a) and (b) in Figure \ref{fig:4loops} arise from the insertion of 4 bi-scalars, which is the maximum at 4 loops, the others are corrections to the 2-loop insertion of 2 bi-scalars. Diagrams with an odd number of such insertions vanish by the tracelessness of the coupling matrices $M$ in (\ref{eq:Mmatrices}) and of products of odd numbers of them. Diagrams with no bi-scalar insertions do not contribute to the $\theta$-Bremsstrahlung as mentioned above. In Figure \ref{fig:4loops} grey bullets represent 1-loop corrections to the gauge propagators whereas the grey box in diagram (p) stands for the internal corrections to the scalar bubble, which we list explicitly in appendix \ref{app:vertex}. 

After selecting the diagrams, the next step consists in the derivation of their algebraic expressions from the ABJM Feynman rules in appendix \ref{app:FeynmanRules}. 
Technically, evaluating the various diagrams involves a bit of index algebra.
We use identities in appendix \ref{app:spinors} to reduce the expressions. We perform the relevant tensor algebra strictly in three dimensions and deal with cumbersome combinations of $\gamma$ matrices in an automated manner with a computer program. In this process we drop all the terms containing an odd number of Levi-Civita tensors, following the remarks above, and reduce all the products of an even number of them to combinations of metric tensors. Finally we obtain expressions featuring scalar products of external velocities $\dot x_i$ and derivatives only. The last step consists in integrating over internal vertices and over the Wilson loop parameters.

\subsection{Momentum integrals and the HQET formalism}\label{sec:HQET}

The greatest simplification granted by setting $\varphi=0$ consists in the fact that integrals arising from Feynman diagrams reduce to 2-point function contributions, instead of retaining a full dependence on the cusp angle. 
This is they evaluate to numbers rather than functions, which makes their evaluation much easier.  

The evaluation of the relevant integrals could in principle be performed directly in $x$-space where the diagrams where computed. This entails solving internal integrations first and then integrating over the Wilson line parameters. The fact that the integrals are propagator-type allows one to employ the powerful Gegenbauer polynomial $x$--space technique GPXT \cite{Chetyrkin:1980pr}. However this approach becomes quite involved for contributions with more than one internal integration, many of which appear in our four loop computation.

Instead, we apply here another strategy, which has proven extremely effective in this sort of settings. This consists in Fourier transforming the integrals to momentum space and perform the Wilson loop contour integrations first, before the integrals over loop momenta, instead of the other way round.
With this procedure computing the integrals boils down to evaluating non-relativistic Feynman integrals of the kind emerging in the context of the heavy quarks effective theory (HQET) (see \cite{Grozin:2014hna,Grozin:2015kna} for related applications in four dimensions). We will  rather use  a Euclidean version of the formalism, since we are going to perform the  computation with $(+,+,+)$ signature.
\begin{figure}[ht]
\begin{center}
\hspace{0.4cm}\includegraphics[width=0.99\textwidth]{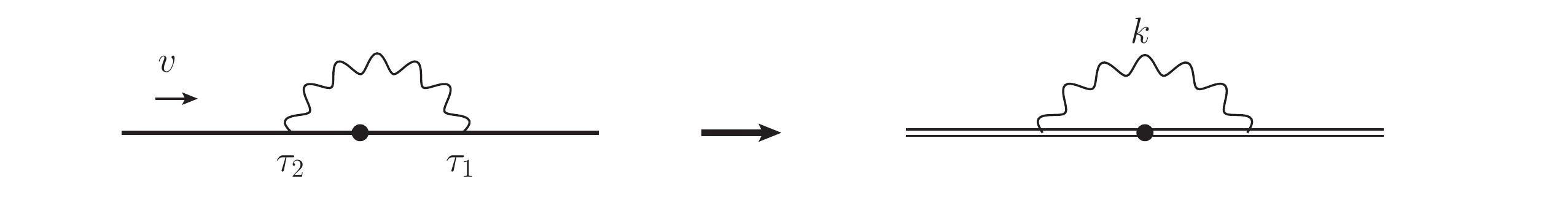}
\begin{equation}
\hspace{-1.5cm} \left[\frac{\G(\frac{1}{2}-\e)}{4 \pi^{3/2-\e}} \right]^2 \!\! \int_{0}^{+\infty}\!\!\!\! d\tau_1\, \int_{-\infty}^{0}\!\! d\tau_2\, \frac{1}{\left[(x_1-x_2)^2\right]^{1/2-\epsilon}} 
\quad \longrightarrow \quad 
\int\, \frac{d^{3-2\epsilon}\,k}{(2\pi)^{3-2\epsilon}}\, \frac{1}{k^2\, (-i\, k\cdot v)^2} \nonumber
\end{equation}
\caption{A cartoon of the Fourier transform of Wilson line integrals to a HQET propagators.}
\label{fig:HQET}
\end{center}
\end{figure}
When applied to a single propagator one-loop case, the Fourier transform is shown pictorially in Figure \ref{fig:HQET}, where the velocity of the heavy quark is the vector tangent to the Wilson loop contour, $v=\dot x(\tau)$. We repeat this also for more complicated loop diagrams appearing in our two- and four-loop computation, obtaining multi-loop HQET integrals.

In general, the resulting HQET integrals suffer from both IR and UV divergences. 
UV divergences are regulated within the framework of dimensional regularization, that is we define space-time integrations in $d=3-2\epsilon$ dimensions. As a consequence, the integrals evaluate to Laurent series in the regularization parameter $\epsilon$, rather to just transcendental numbers. We work in the setting of the dimensional reduction scheme (DRED) \cite{Siegel:1979wq}, which has proven to  be consistent with supersymmetry  (see e.g. 
\cite{Bianchi:2013rma,Griguolo:2013sma,Bianchi:2013pva} for a discussion in the context of WLs in three dimensions). This requires the tensor index algebra in numerators to be performed in strictly three dimensions.

 
IR divergences may also arise from the region of integration at infinity along the Wilson line contour. Following \cite{Grozin:2015kna} we regulate them introducing an exponential factor $e^{\delta\, \tau}$ ($Re(\delta)<0$) for the more external parameters in the path ordered integration. Such a factor suppresses IR effects and enforces the finiteness of the corresponding integrals at large radius. From the HQET standpoint this corresponds to a residual energy for the heavy massive probes, which offsets the HQET propagators by
\begin{equation}\label{eq:IRreg}
\frac{1}{-i\, k\cdot v} \longrightarrow \frac{1}{-i\, k\cdot v - \delta}
\end{equation}
The result of the computation is independent of such a parameter and we conveniently set it to $\delta= -1/2$, a choice that turns out to simplify the relevant integrals.

In conclusions, we Fourier transform all contributions from the diagrams of Figure \ref{fig:4loops}, turning them into heavy quark effective theory (HQET) momentum integrals \cite{Grozin:2015kna,Bianchi:2017svd} and regulate their divergences. The starting strings of the diagrams in momentum space are listed in appendix \ref{app:startingstrings}.

The main advantage of this picture, arises form the possibility of reduction to master integrals, which we spell out in the next section.

Before doing this we observe that the reduction of cusp loop integrals to non-relativistic heavy particles ones has also a very suggestive physical interpretation, beyond its practical function. 
Indeed, the anomalous dimension of the cusped Wilson loop translates in this setting to the renormalization of the current of a massive quark which passes from  velocity $v_1$ to $v_2$ forming an angle $\varphi$.  According to this physical description, the BPS Wilson lines in ABJM theory are associated to heavy W-bosons, transforming in the fundamental representations of the gauge groups and hence interpreted as a massive quark. This particles emerge for instance by Higgsing the theory, moving it away from the origin of the moduli space \cite{Lee:2010hk,Lietti:2017gtc}.

Then, the Bremsstrahlung function associated to the geometric angle $\varphi$ is interpreted as governing the energy loss by radiation of these massive particles undergoing a deviation in its trajectory by an infinitesimal angle $\varphi$.
Analogously, the $\theta$-Bremsstrahlung function which we study in this paper is mapped in the HQET setting to the equivalent, but technically simpler, picture  of a heavy probe with an internal degree of freedom (R-symmetry) undergoing a sudden and infinitesimal kick in internal space, at fixed and vanishing geometrical angle.

\subsection{Master integrals}

The subsequent step involves the  explicit evaluation of the (potentially divergent) Feynman integrals. 
Since all tensor contractions were already performed at the stage of the evaluation in configuration space, one can easily turn all the involved integrals into scalar ones, by rewriting scalar products in terms of inverse propagators.
This leads to integrals with several numerators.
The power of having turned them to momentum HQET integrals stems from the fact that in this form they are amenable of the powerful technique of reduction to master integrals.
In order to perform such a task we repeatedly make use of integration by parts identities (IBP) as shown in the seminal papers \cite{Tkachov:1981wb,Chetyrkin:1981qh}. In practice this step can be cumbersome and an automated implementation is needed to carry it out. In particular, we have used standard software such as LiteRed \cite{Lee:2012cn,Lee:2013mka} and FIRE \cite{Smirnov:2008iw,Smirnov:2013dia,Smirnov:2014hma} to perform this step.
Thanks to the $\varphi=0$ condition, the integrals involved in the computation of the $\theta$-Bremsstrahlung function are precisely those of the kind contributing to the self-energy corrections of a heavy quark. The presence of the cusp point on the line induces only the simple effect of increasing the power of a HQET propagator in the diagram. This occurrence is then dealt with automatically using integration by parts identities which can then be employed to reduce the power of the doubled propagator to unity. 

The outcome of such an analysis is that each diagram can be written as a linear combination on a basis of 21 master integrals. They are sketched in Figure \ref{fig:master} and explicitly defined in appendix \ref{app:masters}.
\begin{figure}
\begin{center}
\includegraphics[scale=0.4]{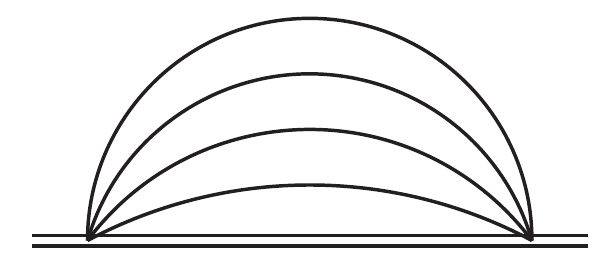} \hspace{2mm}
\includegraphics[scale=0.4]{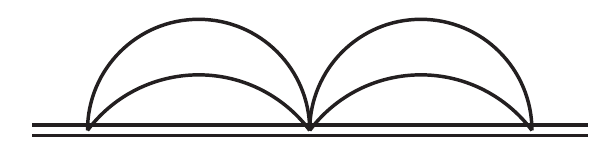} \hspace{2mm}
\includegraphics[scale=0.4]{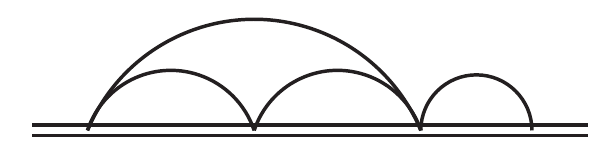} \hspace{2mm}
\includegraphics[scale=0.4]{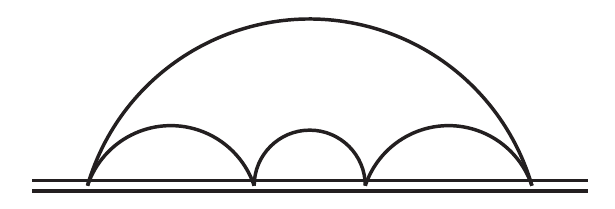} \\[2mm]
\includegraphics[scale=0.4]{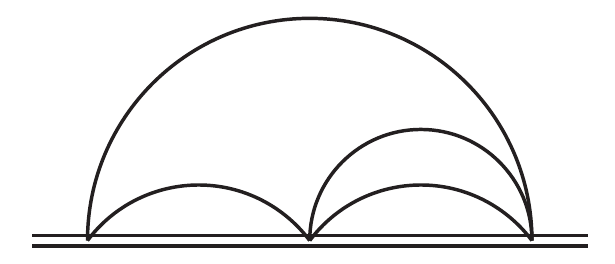} \hspace{2mm}
\includegraphics[scale=0.4]{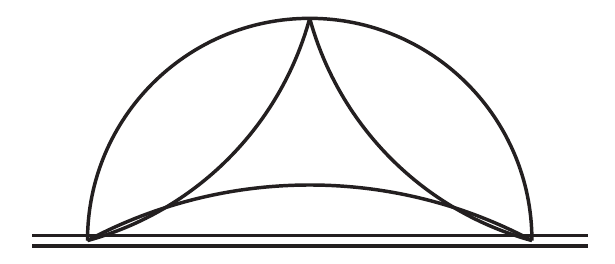} \hspace{2mm}
\includegraphics[scale=0.4]{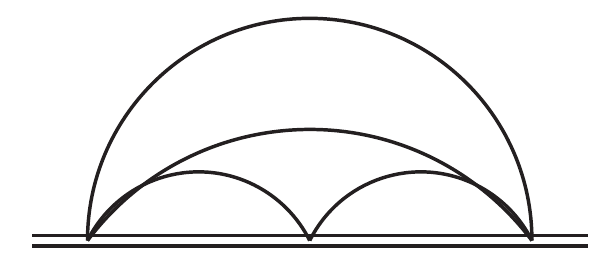} \hspace{2mm}
\includegraphics[scale=0.4]{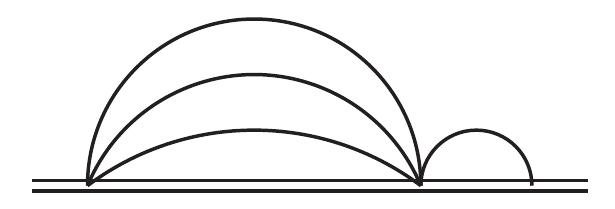} \\[2mm]
\includegraphics[scale=0.4]{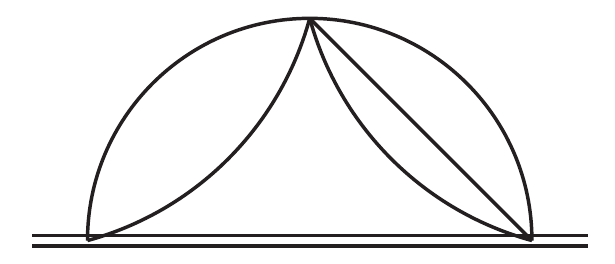} \hspace{2mm}
\includegraphics[scale=0.4]{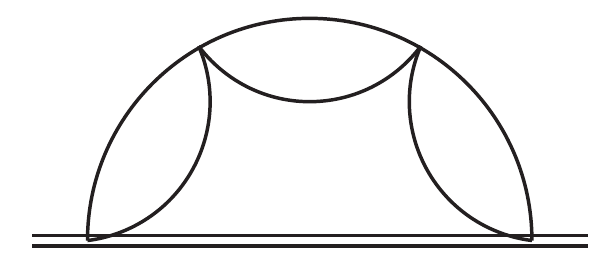} \hspace{2mm}
\includegraphics[scale=0.4]{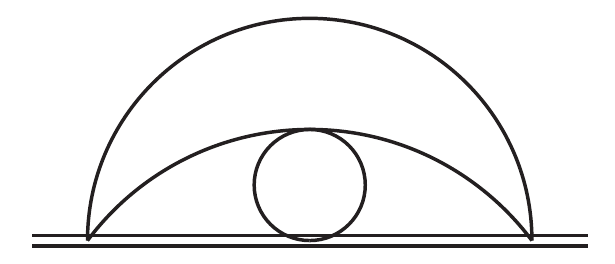} \hspace{2mm}
\includegraphics[scale=0.4]{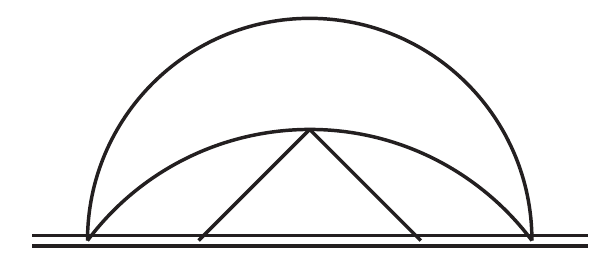} \\[2mm]
\includegraphics[scale=0.4]{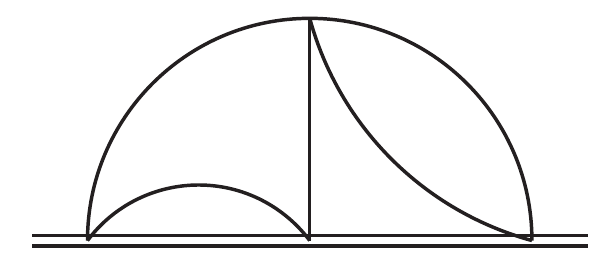} \hspace{2mm}
\includegraphics[scale=0.4]{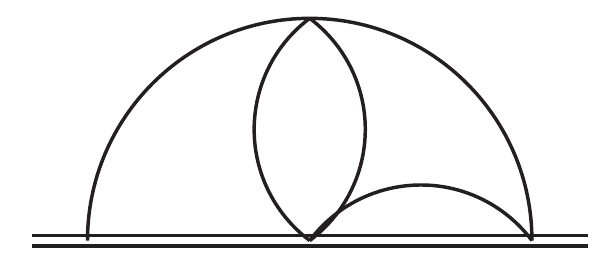} \hspace{2mm}
\includegraphics[scale=0.4]{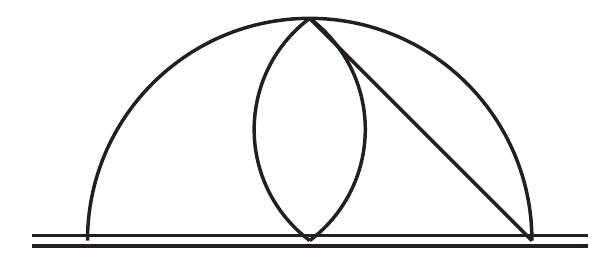} \hspace{2mm}
\includegraphics[scale=0.4]{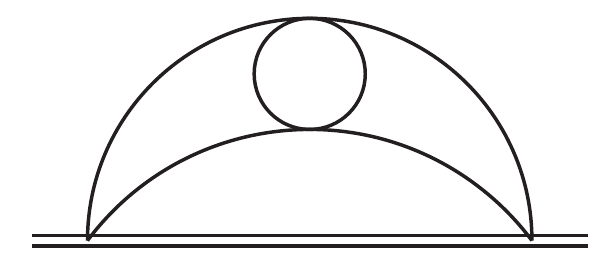} \\[2mm]
\includegraphics[scale=0.4]{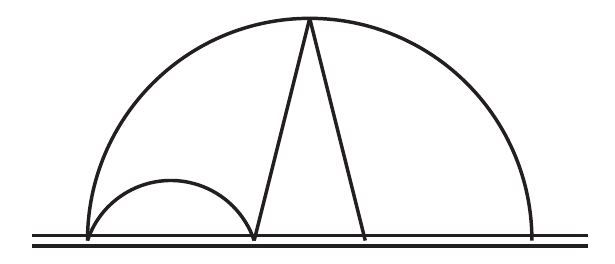} \hspace{2mm}
\includegraphics[scale=0.4]{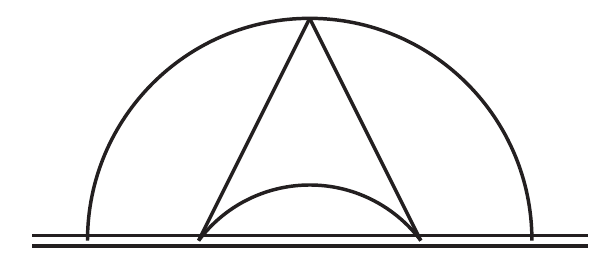} \hspace{2mm}
\includegraphics[scale=0.4]{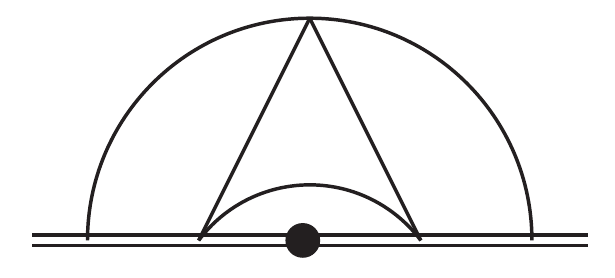} \hspace{2mm}
\includegraphics[scale=0.4]{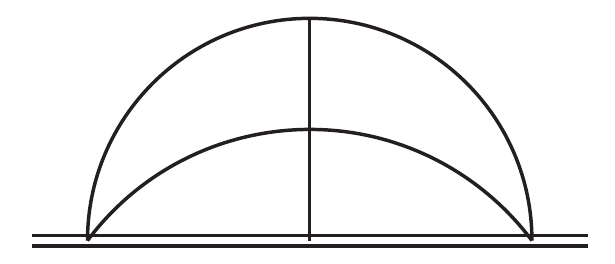} \\[2mm]
\includegraphics[scale=0.4]{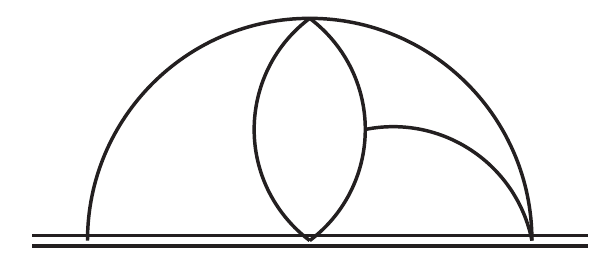}
\end{center}
\caption{Master integrals needed for the computation. The double line represents an HQET propagator. The dot indicates a squared propagator.}\label{fig:master}
\end{figure}
The final step consists in evaluating the master integrals in an $\epsilon$ expansion up to the required order, so as to guarantee a consistent expansion of the cusp expectation value up to the $\frac{1}{\epsilon}$ order. This evaluation is also dealt with in appendix \ref{app:masters}. Using the master integral expressions we eventually find the  $\epsilon$ expansion of the single diagrams of Figure \ref{fig:4loops}, which are collected in appendix \ref{app:diagrams}.

\subsection{The cusp anomalous dimension}\label{sec:detailscusp}

In the previous subsection we detailed the strategy which enables us to perform the four-loop evaluation of the expectation value of the 1/6-BPS Wilson loop on a cusped contour, in the flat cusp limit. 
This expectation value is ultraviolet divergent, due to the cusp. In this section we provide further details and explanation on the renormalization of this object.
We focus on the extraction of the cusp anomalous dimension and its small angle limit, which provides the Bremsstrahlung function.

As recalled above, the expectation value of the cusped Wilson loop possesses both UV and IR divergences. 
We already discussed the introduction of the IR regulator.  UV divergences need further explanations, since they determine the renormalization properties of the Wilson loop and constitute the crucial object of our investigation.   These can in principle originate from different sources.  
At first,  since the ABJM model is conformal, we don't need to consider renormalization of the Lagrangian of the theory. However, divergences associated to the short distance dynamics on the Wilson line can contribute. In the HQET picture these are sourced by the potentially divergent radiative corrections to the heavy quark self-energy. In the case at hand, where the Wilson line is supersymmetric, such divergent contributions are also absent. Finally, a singular geometry of contour induces further divergent contributions. This is precisely the case for a contour with a cusp and this kind of divergence is precisely the one we are interested  in this paper.

The renormalization of non-local operators was studied in a systematic manner in \cite{Dorn:1986dt,Korchemsky:1987wg}. Here we review some basic concepts that we need for our computation. 
\begin{figure}[ht]
\begin{center}
\includegraphics[width=0.75\textwidth]{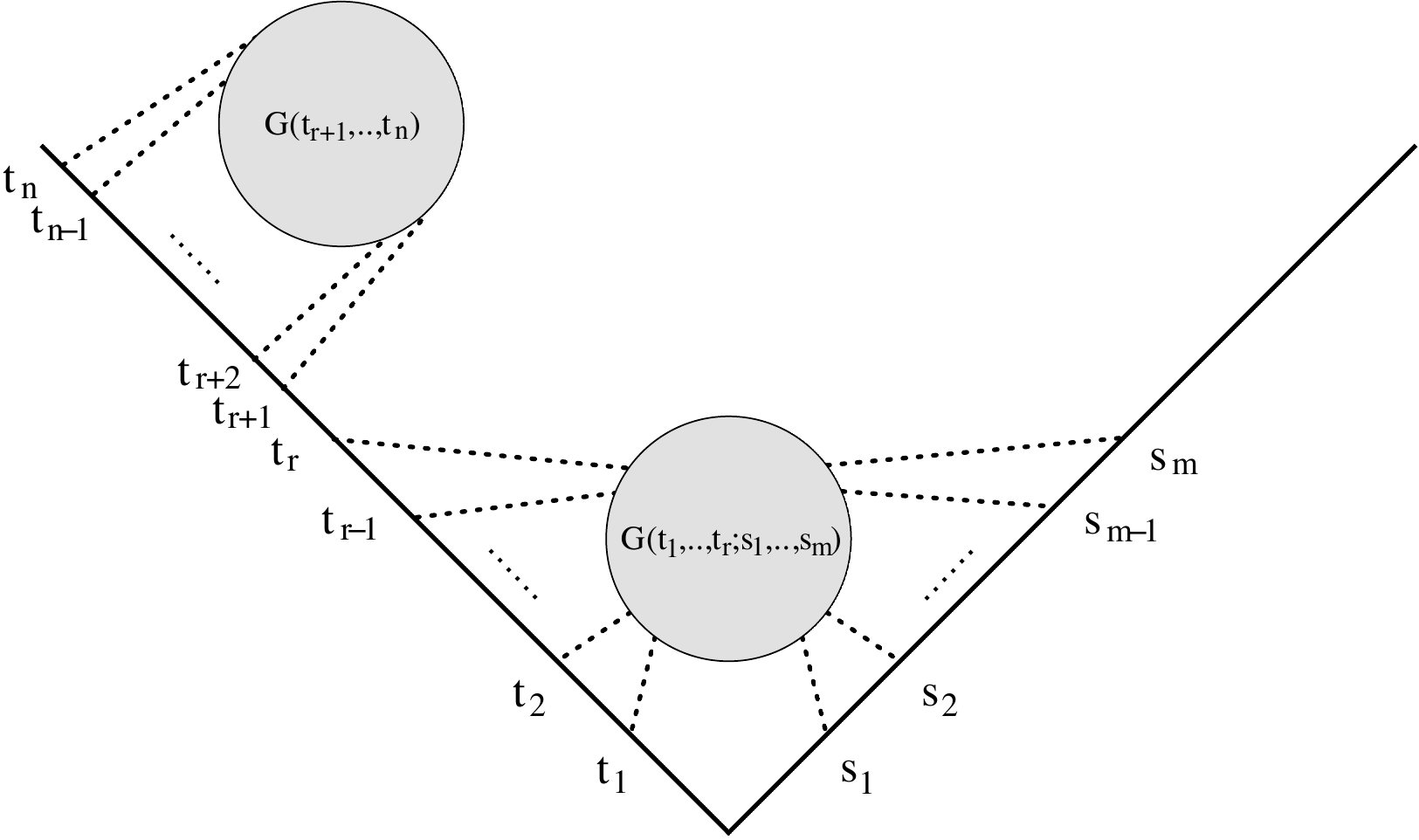}
\caption{Factorization of quantum corrections to the cusped WL.}
\label{fig:factor}
\end{center}
\end{figure}  
We analyze first the contributions which are 1PI vertex diagrams in HQET picture. 
That is, when we consider a configuration like that represented in Figure \ref{fig:factor}, the sub-sector governed by the Green function $G(t_{r+1}, \cdot\cdot, t_{n})$ decouples from the rest. 
Indeed, as discussed in details in \cite{Bianchi:2017svd},  the  contribution of such diagrams can be factorized as the product of a 1PI term which encodes the diagram obtained from the one in Figure \ref{fig:factor} removing the sub-sector  controlled by $G(t_{r+1}, \cdot\cdot, t_{n})$, times a factor  representing the contribution due to $G(t_{r+1}, \cdot\cdot, t_{n})$ to the vacuum expectation value of a straight semi-infinite line running from $-\infty$ to $0$.


Thanks to this factorization property, we can express the cusped Wilson loop expectation value as the sum of all 1PI diagrams, which we denote  by $V(\theta,\varphi)$, times the vacuum expectation value of the semi-infinite Wilson lines $[S(-\infty,0),S(0,\infty)]$, running from $-\infty$ to $0$ and from $0$ to $\infty$, which constitute the rays the cusp is constructed with
\begin{equation}
\label{factorization2}
\langle W(\theta,\varphi)\rangle=S(-\infty,0)V(\theta,\varphi) S(0,\infty)
\end{equation}
In the HQET description these translate into the two-point functions of the heavy quark and \eqref{factorization2} is interpreted as the ordinary decomposition of a correlation function in terms of its 1PI sector and self-energy part. 

In order to single out the cusp anomalous dimension from $\langle W(\theta,\varphi)\rangle$, one should in general perform a subtraction of the self-energy contributions. In practice this amounts to subtracting the contribution of the straight line or the cusp at $\theta=\varphi=0$, leading to the prescription
\begin{equation}
\label{eq:KorchRad2}
\log(\widetilde{W} (\theta,\varphi))=\log\frac{W(\theta,\varphi)}{W(0,0)}=\log \frac{V(\theta,\varphi)}{V(0,0)}
\end{equation}
with $V(\theta,\varphi)$ defined in (\ref{factorization2}). 
In our case, since the contribution of the line is not divergent, and since we are eventually only interested in contributions depending non-trivially on $\theta$, these subtleties can be consistently ignored and one can directly work at the level of the 1PI diagrams, as stated in section \ref{sec:computation}.

Equipped with this formalism, we finally renormalize the UV divergent cusped Wilson loop operator
\begin{equation} \label{eq:WLren}
\langle W_R(\theta,\varphi) \rangle = Z_{1/6}^{-1}\, \langle  \widetilde{W}(\theta,\varphi) \rangle
\end{equation}
From the renormalization constant we extract the cusp anomalous dimension
\begin{equation} \label{eq:cusp16}
\Gamma_{1/6}(k,N_1,N_2) = \frac{d \log Z_{1/6}}{d \log \mu}
\end{equation}
where $\mu$ stems for the renormalization scale, which on dimensional grounds appears at each perturbative loop order $l$ with a power $\mu^{2l}$.
The perturbative computation outlined in section \ref{sec:computation} provides $V(\theta,\varphi)$ as an expansion in $\frac{1}{k^2}$, since as recalled, only even perturbative orders are divergent
\begin{equation}\label{eq:Vexpansion1}
V(\theta,\varphi)=  \left(\frac{2\pi}{k}\right)^2 V^{(2)}(\theta,\varphi) + \left(\frac{2\pi}{k}\right)^4 V^{(4)}(\theta,\varphi) + {\cal O}\left(k^{-6}\right)
\end{equation}
where within dimensional regularization each coefficient $V^{(i)}$ is expressed as a Laurent series in the regularization parameter $\epsilon$.  According to the standard text-book prescription, the cusp anomalous dimension is then extracted from the residues of the simple poles in $\epsilon$ of $Z_{cusp}$, leading to
\begin{equation}\label{eq:cusp}
\log Z_{cusp} = \left. \log\left(\frac{V(\theta,\varphi)}{V(0,0)}\right)\right|_{\frac{1}{\epsilon}\rm terms}\!\!\!\!= - \frac{1}{4\epsilon\, k^2}\, \Gamma^{(2)} - \frac{1}{8\epsilon\, k^4}\, \Gamma^{(4)} + {\cal O}\left(k^{-6}\right)
\end{equation}
Finally, we compute the $\theta$-Bremsstrahlung function by taking the double derivative
\begin{equation}\label{prescription2}
B_{1/6}^{\theta} = \frac12\, \frac{\partial^2}{\partial \theta^2}\, \Gamma_{1/6} \Big|_{\varphi=\theta=0}
\end{equation}
Summing over different diagrams we were able to compute the 4-loop expectation value of the $\theta$-cusped Wilson loop at $\varphi=0$, whose full result we spell out in the next section.

\section{The result}\label{sec:result}

\subsection{Review two-loop result}\label{se:twoloop}

We review here the two--loop computation of the cusped WL expectation value. As detailed above, the two-loop result is needed because the cusp anomalous dimension is extracted from the divergent part of the perturbative logarithm of the full expectation value. Since at four loops we are interested in the order $1/\epsilon$ of the expectation value and its logarithm,  we need to consider the two--loop corrections up to finite terms in the regulator. Moreover, the two loop contributions turn out to be simple examples to describe our computational setting. 

The relevant diagrams are the ones introduced in Figure \ref{fig:2loops}. Parametrizing the points on the cusp line as $x^{\mu}_i(s)= v^{\mu} \t_i$ and using the Feynman rules in appendix \ref{app:FeynmanRules}, the algebra of the first diagram in configuration space gives
\begin{align}
(a) & = N_1 N_2 \left( \frac{2\pi }{k}\right)^2  \left[\frac{\G(\frac{1}{2}-\e)}{4 \pi^{3/2-\e}} \right]^2 \textrm{Tr}(M_1 M_2)   \int^{\infty}_0 d \t_1 \int_{-\infty}^0 d \t_2    \,  \frac{1}{(x_{12}^2)^{1-2\e}} \non
\end{align}
with $x_{12}^2 = (x_1(\t_1)-x_2(\t_2))^2$. This can be Fourier transformed to momentum space using \eqref{eq:Fourier} and, introducing the IR regulator $\delta$, we get
\begin{align}
(a) &  =  4 N_1 N_2 \left( \frac{2\pi }{k}\right)^2  C^2_\theta  \int \frac{d^{3-2\e}k_{1}}{(2\pi)^{3-2\e}}\frac{d^{3-2\e}k_{2}}{(2\pi)^{3-2\e}}  \, \frac{1}{k_2^2(k_1-k_2)^2 (i k_1\cdot v + \delta)^2}  
\end{align}
where  the factor $C_\theta = \cos\frac{\theta}{2}$ is produced by \eqref{eq:traces}. We choose $\delta=-1/2$ and absorb the imaginary unit in the HQET propagator into the velocity $v=i\, \tilde{v}$. The resulting  vector is such that $\tilde{v}^2=-1$ and it can be conveniently used to define the master integrals in euclidean space, making them manifestly real (see appendix \ref{app:masters}). At this stage of the computation the momentum integrals are elaborated by FIRE and projected to the master integral basis.  In the present case the result of integration by parts is rather trivial and we get
\begin{align}
(a) &  = \left( \frac{2\pi }{k}\right)^2   16 N_1 N_2 C^2_\theta (5 - 2 d) G_{0, 1, 1, 0, 1}  
\end{align}
where the master integral $G_{0, 1, 1, 0, 1}$ is defined in \eqref{eq:masterintegrals}. Now the master integral(s) must be evaluated, obtaining the final result for the diagram as an $\e$-expansion up to the desired order
\begin{align}
(a)& =   \left( \frac{2\pi }{k}\right)^2   N_1 N_2 C^2_\theta \bigg(\frac{4 \pi}{\e}+ {\cal O}\left(\epsilon \right) \bigg)
\end{align}
up to an overall factor $4^{2\e} e^{-2 \gamma_E \e}/(4\pi)^{3-2\e}$, omitted to keep the expression compact.
The same procedure can be applied to the second diagram of Figure  \ref{fig:2loops}, which includes all the one loop corrections to the gauge propagator. The final result reads
\begin{align}
(b)& = \left( \frac{2\pi }{k}\right)^2   8 N_1 N_2 \bigg(-\frac{\pi}{\e}+3 \pi+ {\cal O}\left(\epsilon \right) \bigg)
\end{align}
The result of  diagram  (b) does not depend on the angle $\theta $, which means that it does not contribute to the two-loop Bremsstrahlung function $B_{1/6}^{\theta}$. Nevertheless, we need to include its contribution in  $V^{(2)}(\theta)$ in order to consistently extract the perturbative logarithm of the cusp at four-loop order.  Combining the two diagrams we find  
\begin{equation}\label{eq:V1}
V^{(2)}(\theta) =   4\pi  N_1 N_2 \bigg(\frac{C^2_\theta-2}{  \epsilon }+6 +{\cal O}\left(\epsilon \right) \bigg)
\end{equation}
where again an overall factor  $4^{2\e} e^{-2 \gamma_E \e}/(4\pi)^{3-2\e}$ is understood and the dependence on the coupling constant has been stripped out according to \eqref{eq:Vexpansion1}.

\subsection{The four-loop result}

Following the steps outlined in the previous section, we consider the four loop diagrams of Figure \ref{fig:4loops} and express them in momentum space using the rules in appendix \ref{app:FeynmanRules}. We collect the list of starting strings in momentum space in appendix \ref{app:startingstrings}. After gamma algebra manipulation, IBP reduction and evaluating the master integrals we get the $\e$-expansions of the diagrams, collected in appendix \ref{app:diagrams}. 

Putting everything together and using \eqref{eq:KorchRad2} the perturbative computation yields 
\begin{align}\label{eq:W}
\log W\, \Big|_{ \begin{minipage}{10mm} \scriptsize
$\varphi=0$\\ 
$\theta$-dep
\end{minipage}}
\!=
\frac{C_\theta^2 N_1 N_2}{4 k^2 \epsilon }-\frac{C_\theta^2 N_1 N_2^2 \left(\left(6 C_\theta^2+5 \pi ^2-12\right) N_1+\pi ^2 N_2\right)}{48 k^4 \epsilon }+{\cal O}\left(k^{-6}\right) + {\cal O}\left(\epsilon^0\right)
\end{align}
where $C_\theta = \cos\frac{\theta}{2}$. The fact that the logarithm is expressed in terms of a simple pole only already provides a consistency check on the exponentiation of the divergences. In the intermediate steps of the computation, poles in $\epsilon$ up to order 3 are generated in the four loops  1PI expectation value. The cubic order poles are produced by diagrams $(b)$, $(g)$ and $(h)$ (see the expansions in appendix \ref{app:diagrams}), and consistently cancel out. Moreover the coefficient of the remaining double pole is such that it exponentiates the 2-loop result.

One further check is as follows. The one-loop gluon self-energy contains a non-gauge covariant term, which is expected to drop out in physical quantities (see e.g. discussions in \cite{Drukker:2008zx,Griguolo:2012iq}). In fact at 2 loops it can be seen to give rise to an angle independent divergence which is then removed automatically from the gauge invariant cusp anomalous dimension by following the prescription \eqref{eq:KorchRad}. In the 4-loop computation we explicitly kept track of terms arising from this piece and verified that they consistently drop out of the final result \eqref{eq:W}.

Using  \eqref{eq:WLren} ,\eqref{eq:cusp16} and \eqref{prescription2}  we obtain the final result for the $\theta$-Bremsstrahlung function associated with the 1/6-BPS cusp 
\begin{equation}\label{eq:bremsstrahlung}
B_{1/6}^{\theta}(k,N_1,N_2) = \frac{N_1 N_2}{4 k^2}-\frac{\pi ^2 N_1 N_2^2 \left(5 N_1+N_2\right)}{24 k^4}+{\cal O}\left(k^{-6}\right)
\end{equation}
for generic ranks of the gauge groups.
In the ABJM limit of equal ranks this reduces to
\begin{equation}\label{eq:bremsstrahlungABJM}
B_{1/6}^{\theta}(k,N) = \frac{N^2}{4 k^2}-\frac{\pi ^2 N^4}{4 k^4}+{\cal O}\left(k^{-6}\right)
\end{equation}
We notice that the result displays maximal transcendentality (though \eqref{eq:W} does not) and does not contain factors of $\log 2$. It is therefore possible that the $\theta$-Bremsstrahlung has a perturbative expansion in terms of even powers of $\pi$ only, as it appears to be the case for the $\varphi$-Bremsstrahlung.

\section{Comparison with $B_{1/6}^{\varphi}$ and connection to matrix model}\label{sec:conj}

Curiously, the four-loop coefficient displays the same ratio with the two-loop one, as in the conjectured exact $\varphi$-Bremsstrahlung function of \cite{Lewkowycz:2013laa}
\begin{equation}
B_{1/6}^{\varphi}(k,N) = \frac{N^2}{2 k^2} - \frac{\pi^2\, N^4}{2 k^4} +{\cal O}\left(k^{-6}\right) 
\end{equation}
As the cusp does not satisfy a BPS condition for $\varphi=\theta$, the Bremsstrahlung functions associated to the two angles differ  (and indeed they do so already at 2 loops). Nevertheless it is still conceivable that the small angle limits of the cusp anomalous dimension are related in a simple fashion.
From our four loop result it would be tempting to extrapolate an all order relation 
\begin{equation}
B_{1/6}^{\varphi}(k,N) \underset{\text{conj}}{=} 2\, B_{1/6}^{\theta}(k,N)
\end{equation}
though this is a quite bold statement at this stage. A confirmation or disproof could come for instance by a strong coupling computation of $B_{1/6}^{\theta}$, but this is lacking, to the best of our knowledge \cite{Correa:2014aga,Aguilera-Damia:2014bqa}.

We now comment on the color structure of the result \eqref{eq:bremsstrahlung}.
The $N_1 N_2^3$ term was predicted in \cite{Bianchi:2016rub}, as part of an all-order computation of the terms associated to the highest $N_2$ power $N_2^{l-1}$ at a given perturbative order $l$.
The result presented here is in complete agreement with this prediction and confirms it.
The $N_1^2 N_2^2$ term is new as is the $N_1^3 N_2$ contributions that happens to vanish, as a result of remarkable cancellations across different diagrams. We do not have a particular insight on this fact.

Still, we point out the following remarkable fact on the color structure of the result \eqref{eq:bremsstrahlung}.
The conjecture \cite{Lewkowycz:2013laa} on the exact $B_{1/6}^{\varphi}$ function was derived in ABJM theory, that is with equal ranks, by relating it to a circular multiple wound 1/6-BPS Wilson loop
\begin{equation}\label{eq:exactB}
B_{1/6}^{\varphi}(k,N) = \frac{1}{4\pi^2}\, \partial_n\, |W_n(k,N)|\, \bigg|_{n=1}
\end{equation}
The Wilson loop can be computed exactly from localization \cite{Klemm:2012ii} and its result reads
\begin{align}
\langle W_n \rangle(k,N) &= 1 + i \pi   n^2  \frac{N}{k}  +  \left(\frac{2 \pi ^2 n^2}{3}  -\frac{\pi ^2 n^4}{3}\right) \frac{N^2}{k^2}  -\frac{N^3}{18k^3} i \pi ^3   n^2 \left(n^4-8 n^2+4\right) 
\nonumber\\&
+\frac{\pi ^4 N^4}{180k^4} n^2 \left(n^6-20 n^4+58 n^2-60\right) +{\cal O}\left(k^{-5}\right)
\end{align}
We can now consider the same Wilson loop in the ABJ model. Expanding the matrix model of \cite{Marino:2009jd} in this case we obtain the expectation value (the expression up to order 8 can be found in the appendices of \cite{Bianchi:2016gpg})
\begin{align}\label{eq:WLn}
\langle W_n \rangle(k,N_1,N_2) &= 1 + \frac{i \pi  n^2 N_1}{k} 
-\frac{\pi ^2 n^2 N_1 \left(n^2 N_1+N_1-3 N_2\right)}{3 k^2}  \nonumber\\&
-\frac{i \pi ^3 n^2 N_1 \left(-6 N_1 \left(2 n^2 N_2+N_2\right)+\left(n^4+4 n^2+1\right) N_1^2+9 N_2^2\right)}{18 k^3}  \nonumber\\& +
\frac{\pi ^4 n^2 N_1}{180 k^4} \left(n^6 N_1^3+10 n^4 N_1^2 \left(N_1-3 N_2\right) \right. \nonumber\\& \left. + n^2 N_1 \left(13 N_1^2-75 N_2 N_1+120 N_2^2\right) -30 N_2^2 \left(N_1+N_2\right)\right) + \dots
\end{align}
If we plug this expression into \eqref{eq:exactB}, even though we do not have a proof that this gives the $B_{1/6}^{\varphi}$-Bremsstrahlung also in the ABJ case, we obtain
\begin{align}\label{eq:conjecture}
B_{1/6}^{\varphi}(k,N_1,N_2) &\underset{\text{conj}}{=} 2\, B_{1/6}^{\theta}(k,N_1,N_2) \underset{\text{conj}}{=} \frac{1}{4\pi^2}\, \partial_n\, |W_n(k,N_1,N_2)|\, \bigg|_{n=1}  \nonumber\\& = \frac{N_1 N_2}{2 k^2}-\frac{\pi ^2 N_1 N_2^2 \left(5 N_1+N_2\right)}{12 k^4}  \nonumber\\&
+ \frac{\pi ^4 N_1 N_2^2 \left(-23 N_1^3+345 N_2 N_1^2+145 N_2^2 N_1+3 N_2^3\right)}{720 k^6}
 \nonumber\\&
-\frac{\pi ^6 N_1 N_2^2}{30240 k^8} \left(95 N_1^5-2331 N_2 N_1^4+17633 N_2^2 N_1^3 \right.\nonumber\\&\phantom{-\frac{\pi ^6 N_1 N_2^2}{30240 k^8}}\left. +12285 N_2^3 N_1^2+875 N_2^4 N_1+3 N_2^5\right) +{\cal O}\left(k^{-10}\right)
\end{align}
Remarkably, the color components of the four-loop result are in the same ratio as in \eqref{eq:bremsstrahlung}.
This could be again only a coincidence, but it seems to hint that \eqref{eq:bremsstrahlung} can be obtained as a derivative of a winding Wilson loop, hence corroborating \eqref{eq:conjecture}.
It would be interesting to extend the calculation presented here to the color subleading corrections and inspect whether a relationship with the corresponding nonplanar piece of the winding Wilson loop still holds (as in \cite{Bianchi:2017svd}).

We further comment on the color structure of the conjectured result \eqref{eq:conjecture}.
At each order we can factorize a common $N_1 N_2^2$. This means that on a possible range of color structures at each order $l$ in perturbation theory $\{ N_1^l N_2^0, N_1^{l-1}N_2^1, \dots, N_1\, N_2^{l-1}, N_1^0 N_2^{l} \}$ only the terms $\{ N_1^{l-2}N_2^2, \dots, N_1\, N_2^{l-1} \}$ appear.
The potential contribution proportional to $N_2^l$ can not be present by construction, whereas the structure $N_1^l$ would correspond to a pure Chern-Simons piece, which, albeit it appears in the circular Wilson loop \eqref{eq:WLn}, is not expected to contribute to the cusp anomalous dimension, and indeed its coefficient vanishes in \eqref{eq:conjecture}. The surprising fact is that also the part proportional to $N_1^{l-1}N_2$ seems to be consistently absent in the proposal \eqref{eq:conjecture} (at least to the perturbative order we probed). We lack an explanation for this phenomenon.

\section{Conclusions}

In this paper we have analyzed the Bremsstrahlung function associated to the internal angle $\theta$ for the locally 1/6-BPS generalized cusp in the ABJM model. We have performed its computation at four loops at weak coupling in the planar limit.

Technically, our computation has considerably benefited from considering only contributions which are relevant for the computation of the Bremsstrahlung function. This in particular allowed the expansion of the cusp anomalous dimension in the small internal angle in the R-symmetry space, at vanishing geometric angle. 
This limit entails remarkable simplifications at both the level of the number of diagrams involved, and of their practical evaluation, especially when dealing with the integrals.
These indeed reduce at $\varphi=0$ to self-energy contributions, which we computed by turning to the HQET picture (via Fourier transform to momentum space) and employing integration by parts identities to reduce them to a restricted set of master topologies.

The aim of our four-loop computation consists in gathering more information on $B_{1/6}^{\theta}$, which is thus far a quite elusive object.
Indeed, we remind that other Bremsstrahlung functions in ABJM, namely that related to the 1/2-BPS cusp and that associated to the geometric angle $\varphi$ of the 1/6-BPS cusp have already been given exact expressions \cite{Lewkowycz:2013laa,Bianchi:2014laa}.
This has been achieved by relating them to the expectation value of supersymmetric 1/6-BPS Wilson loops with multiple winding, which are computable exactly via localization \cite{Klemm:2012ii}.
On the contrary, no such expressions so far had been derived for $B_{1/6}^{\theta}$.

Based on the four-loop result we have proposed a conjecture that relates $B_{1/6}^{\theta}$ to $B_{1/6}^{\varphi}$ and consequently provides an exact expression for the former.
We recall that no prediction for $B_{1/6}^{\theta}$ at string 't Hooft coupling is available from the dual string theory picture on the background $AdS_4\times CP^3$.
Our exact proposal, relating the Bremsstrahlung function to a multiply wound supersymmetric Wilson loop known exactly via localization, allows for formulating such a prediction.
Namely, at strong coupling the Bremsstrahlung function has the expansion (for $N_1=N_2=N\gg 1$)
\begin{equation}
B_{1/6}^{\theta} = \frac{\sqrt{\lambda}}{4\sqrt{2}\pi} - \frac{1}{8\pi^2} + \left( \frac{1}{8\pi^3} + \frac{5}{192\pi} \right) \frac{1}{\sqrt{2\lambda}} + {\cal O}\left(\lambda^{-3/2}\right) \qquad \lambda\equiv \frac{N}{k}\gg 1
\end{equation}

The exact knowledge (albeit still conjectural) of the $\theta$-Bremsstrahlung function is already interesting, being another example of a non-BPS observable which can be computed (though indirectly) with a localization result. 
Moreover the result can also be relevant in view of a potential computation of the same quantity based on integrability, as carried out in ${\cal N}=4$ SYM.
Recent developments on the Quantum Spectral Curve approach \cite{Gromov:2013pga,Gromov:2014caa,Gromov:2015dfa,Gromov:2016rrp} in the ABJM model \cite{Cavaglia:2014exa,Bombardelli:2017vhk} have provided progress in this direction. 
We stress that an integrability based computation would not only provide a non-trivial crossed check of the localization based proposal \eqref{eq:conjecture}, but would also grant a direct proof of the conjecture on the exact expression for the interpolating $h$ function of ABJM \cite{Nishioka:2008gz,Grignani:2008is,Gaiotto:2008cg,Gromov:2014eha}.

In \cite{Cavaglia:2016ide} a proposal appeared on how to relate observables in ABJM and ABJ theories which can be computed via integrability.
In ABJM, integrability based computations are given in terms of the interpolating function $h(\lambda)$ \cite{Nishioka:2008gz, Gaiotto:2008cg, Grignani:2008is}, whose exact expression was conjectured in \cite{Gromov:2014eha}. This matches the perturbative data at weak \cite{Nishioka:2008gz, Grignani:2008is, Minahan:2009aq, Minahan:2009wg, Leoni:2010tb} and strong coupling \cite{McLoughlin:2008he, Abbott:2010yb, LopezArcos:2012gb}. 
Assuming that ABJ is also integrable and according to the prescription of \cite{Cavaglia:2016ide}, the same observable in ABJ would then be obtained by replacing the 't Hooft coupling $\lambda = \frac{N}{k}$ with an effective ABJ version $\lambda_{eff}(N_1,N_2)$, whose explicit expression can be found in \cite{Cavaglia:2016ide}.
Assuming that the $\theta$-Bremsstrahlung computed in this note could indeed be computed via integrability, we however observe that the replacement $\lambda\rightarrow \lambda_{eff}$ described above fails to reproduce \eqref{eq:bremsstrahlung} from \eqref{eq:bremsstrahlungABJM}.
This indicates that some of the assumptions above do not hold in this case or that the prescription of \cite{Cavaglia:2016ide} somehow does not directly apply in this case.
Still, the ABJ theory is expected to be integrable (it was proven to be so in a particular sector in the limit of \cite{Bianchi:2016rub}), therefore a derivation of its Bremsstrahlung function from integrability is also foreseeable. This, together with a deeper understanding of the ABJ supersymmetric cusp, would grant a firmer handle on the conjecture for the exact interpolating function of the ABJ model \cite{Cavaglia:2016ide}.

We conclude with remarks on possible perspectives.
In \cite{Bianchi:2014laa,Correa:2014aga} a connection was conjectured between the $\theta$-Bremsstrahlung and the derivative of deformed circular BPS Wilson loops \cite{Cardinali:2012ru}.
It would be interesting to compare the four-loop computation described here with the expectation value of such a Wilson loop at the same order.
However the crucial simplifications described in section \ref{sec:computation} that made this computation doable are absent in the case of a circular Wilson loop and as a result its evaluation would be rather complicated.
On the contrary, a setting where part of the simplification employed here still applies, would be the computation of the $\varphi$-Bremsstrahlung (a part of the four-loop Bremsstrahlung function in QCD in four dimensions has been recently performed \cite{Grozin:2017css}). In that case one would have to perform derivatives of the cusp with respect to $\varphi$, but eventually setting it to 0 would still allow to use propagator type HQET integrals of the kind used in this paper. However the full computation would require far more diagrams (for instance those using the gluon 2-, 3- and 4-point functions at 3, 2 and 1 loops respectively) and master integrals. It would be interesting to perform such a computation so as to test the conjecture of \cite{Lewkowycz:2013laa} on the exact $\varphi$-Bremsstrahlung function at four loops.

\acknowledgments

We would like to thank L. Griguolo, S. Penati, M. Preti and D. Seminara for helpful discussions.
This work has been supported in part by Italian Ministero dell'Istruzione, Universit\`a e Ricerca (MIUR) and Istituto
Nazionale di Fisica Nucleare (INFN) through the ``Gauge Theories, Strings, Supergravity'' (GSS) research project.

\vfill
\newpage

\appendix

\section{Spinor  and  group conventions}\label{app:spinors}

We work in euclidean three dimensional space with coordinates $x^\mu = (x^0, x^1, x^2)$. The Dirac matrices satisfying the Clifford algebra $\{ \g^\mu , \g^\nu \} = 2 \d^{\mu\nu} \mathbb{I}$ are chosen to be
\beq
\label{diracmatrices}
(\g^\mu)_\a^{\; \, \b} = \{ -\s^3, \s^1, \s^2 \}
\eeq
with matrix product 
\beq
\label{prod}
(\g^\mu \g^\nu)_\a^{\; \, \b} \equiv (\g^\mu)_\a^{\; \, \g} (\g^\nu)_\g^{\; \, \b}
\eeq
The algebra of the matrices \eqref{diracmatrices} is completely determined by the relation
\begin{align}
& \g^\mu \g^\nu = \d^{\mu \nu} \mathds{1} - i \varepsilon^{\mu\nu\rho} \g^\rho
 \end{align}
which gives rise to the traces 
\begin{align}
\Tr (\g^\mu \g^\nu) = 2 \d^{\mu\nu}\  \ \ \ \ \ \ \ \ 
\textrm{and}\  \ \ \ \ \ \ \ \ 
\Tr (\g^\mu \g^\nu \g^\rho) = -2i \varepsilon^{\mu\nu\rho}
\end{align}
Spinor indices are raised and lowered by means of the $\epsilon-$tensor: 
\beq
\psi^\a = \varepsilon^{\a\b} \psi_\b \quad  \quad \psi_\a = \varepsilon_{\a\b}  \psi^\b    
\eeq
with $\varepsilon^{12} = - \varepsilon_{12} = 1$. In particular,  the antisymmetric combination of two spinors  can be reduced to scalar contractions:
\begin{align}
\psi_\a \chi_\b - \psi_\b \chi_\a = \varepsilon_{\a\b} \psi^\g \chi_\g\equiv \varepsilon_{\a\b} \psi\chi,
\ \ \ \ \ \ 
 \psi^\a \chi^\b - \psi^\b \chi^\a = -\varepsilon^{\a\b} \psi^\g \chi_\g\equiv-\varepsilon^{\a\b} \psi \chi
\end{align}
Under complex conjugation the gamma matrices transform as follows: $[(\g^\mu)_\a^{\; \, \b}]^* = (\g^\mu)^\b_{\; \, \a} \equiv \e^{\b\g} (\g^\mu)_\g^{\; \, \d}  \e_{\a\d}$. As a consequence, the hermitian conjugate of the vector bilinear  can be rewritten as follows
\beq
(\psi \g^\mu \chi)^\dagger =(\psi^\a (\g^\mu)_\a^{\; \, \b} \chi_\b)^\dagger = \bar{\chi}_\b (\g^\mu)^\b_{\; \, \a} \bar{\psi}^\a = \bar{\chi}^\b  (\g^\mu)_\b^{\; \, \a} \bar{\psi}_\a \equiv 
\bar{\chi} \g^\mu \bar{\psi}
\eeq 
where we have taken $ (\chi_\beta)^\dagger=\bar\chi_\beta$ and $ (\psi^\alpha)^\dagger=\bar\psi^\alpha$.

\noindent
The U$(N)$  generators are defined as $T^A = (T^0, T^a)$, where $T^0 =
\frac{1}{\sqrt{N}}\mathds{1}$ and $T^a$ ($a=1,\ldots, N^2-1$) are an orthonormal  set of traceless
$N\times N$ hermitian matrices.  The generators are normalized as
\be
\Tr( T^A T^B )= \delta^{AB}
\ee 
The structure constant are then defined by 
\be
[T^A,T^B]=i f^{AB}{}_C T^C
\ee
In the paper  we shall often use  the double notation and  the fields will carry two indices in the fundamental representation
of the gauge groups. An index in the fundamental representation of U$(N_1)$ will be generically by the  lowercase roman indices $i,j,k,\dots$, while for an index  in the fundamental representation of U$(N_2)$ we shall use the hatted lowercase roman indices $\hat i, \hat j, \hat k,\dots$

\section{The ABJM action}\label{app:ABJM}

We summarize here the  basic features of the action for general U$(N_1)_k \times$ U$(N_2)_{-k}$ ABJ(M) theories. The gauge sector contains two gauge fields 
$(A_\mu)_i{~}^j$  and $(\hat {A}_\mu)_{\hat i}{~}^{\hat{j}}$ belonging respectively to
the adjoint of U$(N_1)$ and U$(N_2)$. The matter sector instead consists of  the complex fields
$(C_I)_i{~}^{\hat j}$ and $(\bar {C}^I)_{\hat{i}}{~}^j$ as well as the fermions $(\psi_I)_i{~}^{\hat j}$ and $(\bar {\psi}^I)_{\hat{i}}{~}^j$ . The fields $(C_I, \bar\psi^I)$ transform in 
the $({\bf N_1},{\bf \bar N_2})$ of the gauge group while the couple $(\bar C^I,\psi_I)$ belongs to the
representation $({\bf \bar N_1},{\bf N_2})$.
The additional capital index $I = 1,2,3,4$  belongs to the R--symmetry group $SU(4)$. In
order to quantize the theory at the perturbative level, we  introduce the usual  gauge--fixing  for both gauge fields and the  two corresponding sets of ghosts $(\bar c,c)$ and $(\bar{\hat c},\hat c)$. Then the action contains four different contributions
\beq
\label{action}
S = S_{\mathrm{CS}} \big|_{\mathrm{g.f.}}+ S_{\mathrm{mat}} + S_{\mathrm{pot}}^{\mathrm{bos}} + S_{\mathrm{pot}}^{\mathrm{ferm}} 
\eeq 
where
\begin{subequations}
\begin{align}
\label{action1}
S_{\mathrm{CS}}\big|_{\mathrm{g.f.}} &=\frac{k}{4\pi}\int d^3x\,\varepsilon^{\mu\nu\rho} \Big\{ i \, \Tr \!\left(\hat{A}_\mu\partial_\nu 
\hat{A}_\rho+\frac{2}{3} i \hat{A}_\mu \hat{A}_\nu \hat{A}_\rho \right) \!-\! i \, \Tr \left( A_\mu\partial_\nu A_\rho+\frac{2}{3} i A_\mu A_\nu A_\rho\! \right)\!\non \\
&~   + \, \Tr \Big[ \frac{1}{\xi}  (\pa_\mu A^\mu)^2 -\frac{1}{\xi} ( \pa_\mu \hat{A}^\mu )^2 + \pa_\mu \bar{c} D^\mu c  
  - \pa_\mu \bar{\hat{c}} D^\mu \hat{c} \Big] \Big\}
\\
\label{action2}
S_{\mathrm{mat}} =& \int d^3x \, \Tr \Big[ D_\mu C_I D^\mu \bar{C}^I - i \bar{\Psi}^I \g^\mu D_\mu \Psi_I \Big] 
\\
 S_{\mathrm{pot}}^{\mathrm{bos}} =& -\frac{4\pi^2}{3 k^2} \int d^3x \, \Tr \Big[ C_I \bar{C}^I C_J \bar{C}^J C_K \bar{C}^K + \bar{C}^I C_I \bar{C}^J C_J \bar{C}^K C_K\non
 \\
&~ \qquad \qquad \qquad \qquad + 4 C_I \bar{C}^J C_K \bar{C}^I C_J \bar{C}^K - 6 C_I \bar{C}^J C_J \bar{C}^I C_K \bar{C}^K \Big] 
\\
 S_{\mathrm{pot}}^{\mathrm{ferm}} =&  -\frac{2\pi i}{k} \int d^3x \, \Tr \Big[ \bar{C}^I C_I \Psi_J \bar{\Psi}^J - C_I \bar{C}^I \bar{\Psi}^J \Psi_J
+2 C_I \bar{C}^J \bar{\Psi}^I \Psi_J 
\non \\
&~ \qquad \qquad  - 2 \bar{C}^I C_J \Psi_I \bar{\Psi}^J - \e_{IJKL} \bar{C}^I\bar{\Psi}^J \bar{C}^K \bar{\Psi}^L + \e^{IJKL} C_I \Psi_J C_K \Psi_L \Big]
\end{align}
\end{subequations}
 The invariant $SU(4)$ tensors $\epsilon_{IJKL}$ and $\epsilon^{IJKL}$ satisfy $\e_{1234}=\e^{1234} =1$.   The covariant derivatives are  defined  as
\bea
\label{covariant}
D_\mu C_I &=& \pa_\mu C_I + i A_\mu C_I - i C_I \hat{A}_\mu,
\quad \quad 
D_\mu \bar{C}^I = \pa_\mu \bar{C}^I - i \bar{C}^I A_\mu + i \hat{A}_\mu \bar{C}^I
\non \\
D_\mu \bar{\Psi}^I  &=& \pa_\mu \bar{\Psi}^I + i A_\mu \bar{\Psi}^I - i \bar{\Psi}^I \hat{A}_\mu,
\quad  \quad
D_\mu \Psi_I = \pa_\mu \Psi_I - i \Psi_I A_\mu + i \hat{A}_\mu \Psi_I  
\eea

\section{Feynman rules}\label{app:FeynmanRules}

We use the Fourier transform definition
\begin{equation}\label{eq:Fourier}
\int \frac{d^{3-2\e}p}{(2 \p)^{3-2\e}} \frac{p^{\m}}{(p^2)^s} e^{i p \cdot (x-y)} =  \frac{\G(\frac{3}{2}-s-\e)}{4^s \pi^{3/2-\e}\Gamma(s)} \big(-i \partial^{\m}_x \big)\frac{1}{(x-y)^{2(3/2-s-\e)}}
\end{equation}
In euclidean space we define the functional generator as $Z \sim \int e^{-S}$, with action  \eqref{action}.  This gives rise to the following Feynman rules
\begin{itemize}
\item Vector propagators in Landau gauge
\bea
\label{treevector}
 \langle (A_\mu)_i{}^j (x) (A_\nu)_{k}{}^\ell(y) \rangle^{(0)} &=&  \d^{\ell}_i \delta_k^j  \, \left( \frac{2\pi i}{k} \right) \frac{\G(\frac32-\e)}{2\pi^{\frac32 -\e}} \varepsilon_{\mu\nu\rho} \frac{(x-y)^\rho}{[(x-y)^2]^{\frac32 -\e} }
\non \\
&=&  \d^{\ell}_i \delta_k^j   \left( \frac{2\pi }{k} \right) \varepsilon_{\mu\nu\rho} \, \int \frac{d^np}{(2\pi)^n}   \frac{p^\rho}{p^2} e^{ip(x-y)}
\non \\
\non \\
 \langle (\hat{A}_\mu)_{\hat{i}}{}^{\hat{j}} (x) ( \hat{A}_\nu)_{\hat k}{}^{\hat\ell }(y) \rangle^{(0)} &=&  - \d^{\hat\ell}_{\hat i} \delta_{\hat k}^{\hat j}     \, \left( \frac{2\pi i}{k} \right) \frac{\G(\frac32-\e)}{2\pi^{\frac32 -\e}} \varepsilon_{\mu\nu\rho} \frac{(x-y)^\rho}{[(x-y)^2]^{\frac32 -\e} }
\non \\
&=& -\d^{\hat\ell}_{\hat i} \delta_{\hat k}^{\hat j} \left( \frac{2\pi }{k} \right) \varepsilon_{\mu\nu\rho} \, \int \frac{d^np}{(2\pi)^n}   \frac{p^\rho}{p^2} e^{ip(x-y)}
\eea
\item Scalar propagator
\bea
\label{scalar}
\langle (C_I)_i{}^{\hat{j}} (x) (\bar{C}^J)_{\hat{k}}{}^l(\; y) \rangle^{(0)}  &=& \d_I^J \d_i^l \d_{\hat{k}}^{\hat{j}} \, \frac{\G(\frac12 -\e)}{4\pi^{\frac32-\e}} 
\, \frac{1}{[(x-y)^2]^{\frac12 -\e}}
\non \\
&=& \d_I^J \d_i^l \d_{\hat{k}}^{\hat{j}} \, \int \frac{d^np}{(2\pi)^n}   \frac{e^{ip(x-y)} }{p^2}  
\eea
\item Fermion propagator
\bea
\label{treefermion}
\langle (\psi_I^\a)_{\hat{i}}{}^{ j}  (x) (\bar{\psi}^J_\b )_k{}^{ \hat{l}}(y) \rangle^{(0)} &=&  i \, \d_I^J \d_{\hat{i}}^{\hat{l}} \d_{k}^{j} \, 
\frac{\G(\frac32 - \e)}{2\pi^{\frac32 -\e}} \,  \frac{(\g^\mu)^\a_{\; \, \b} \,  (x-y)_\mu}{[(x-y)^2]^{\frac32 - \e}}
\non \\
&=& \d_I^J \d_{\hat{i}}^{\hat{l}} \d_{k}^{j} \, \int \frac{d^np}{(2\pi)^n}   \frac{(\g^\mu)^\a_{\; \, \b} \, p_\mu  }{p^2}  e^{ip(x-y)}
\eea
\item Gauge cubic vertex
\beq
\label{gaugecubic}
i \frac{k}{12\pi} \varepsilon^{\mu\nu\rho} \int d^3x \, f^{abc} A_\mu^a A_\nu^b A_\rho^c
\eeq
\item Gauge-fermion cubic vertex
\beq
\label{gaugefermion}
-\int d^3x \, \Tr \Big[ \bar{\Psi}^I \g^\mu \Psi_I A_\mu - \bar{\Psi}^I \g^\mu \hat{A}_\mu \Psi_I  \Big]
\eeq 
\end{itemize}

\noindent
The one loop gauge propagators are given by
\begin{subequations}
\begin{align}
\label{1vector}
 \langle (A_\mu)_i{}^j (x) &(A_\nu)_{k}{}^\ell(y) \rangle^{(1)} =  \d^{\ell}_i \delta_k^j    \left( \frac{2\pi }{k} \right)^2 N_2 \, \frac{\G^2(\frac12-\e)}{4\pi^{3 -2\e}} 
\left[ \frac{\d_{\mu\nu}}{ [(x- y)^2]^{1-2\e}} - \pa_\mu \pa_\nu \frac{[(x-y)^2]^{2\e}}{4\e(1+2\e)} \right]  
\non  \\
  =&   \d^{\ell}_i \delta_k^j    \left( \frac{2\pi }{k} \right)^2 N_2 \, \frac{\G^2(\frac12-\e)\G(\frac12 +\e)}{\G(1-2\e) 2^{1-2\e} \pi^{\frac32 -\e}} 
\, \int \frac{d^np}{(2\pi)^n}   \frac{e^{ip(x-y)}}{(p^2)^{\frac12 +\e}}  \left( \d_{\mu \n} - \frac{p_\mu p_\nu}{p^2} \right) 
\\
 \langle (\hat{A}_\mu)_{\hat{i}}{}^{\hat{j}} (x) &( \hat{A}_\nu)_{\hat k}{}^{\hat\ell }(y) \rangle^{(1)} = \d^{\hat\ell}_{\hat i} \delta_{\hat k}^{\hat j}     \left( \frac{2\pi }{k} \right)^2 N_1 \, \frac{\G^2(\frac12-\e)}{4\pi^{3 -2\e}} 
\left[ \frac{\d_{\mu\nu}}{ [(x- y)^2]^{1-2\e}} - \pa_\mu \pa_\nu \frac{[(x-y)^2]^{2\e}}{4\e(1+2\e)} \right] \non \\
 =& \d^{\hat\ell}_{\hat i} \delta_{\hat k}^{\hat j}   \left( \frac{2\pi }{k} \right)^2 N_1 \, \frac{\G^2(\frac12-\e)\G(\frac12 +\e)}{\G(1-2\e) 2^{1-2\e} \pi^{\frac32 -\e}} 
\, \int \frac{d^np}{(2\pi)^n}   \frac{e^{ip(x-y)}}{(p^2)^{\frac12 +\e}} \left( \d_{\mu \n} - \frac{p_\mu p_\nu}{p^2} \right) 
\end{align}
\end{subequations}
The one-loop fermion propagator reads 
\bea
\label{1fermion}
&& \langle (\psi_I^\a)_{\hat{i}}{}^{\; j}  (x) (\bar{\psi}^J_\b)_k{}^{\; \hat{l}}(y) \rangle^{(1)} =  -i \,\left( \frac{2\pi}{k} \right) \,  \d_I^J \d_{\hat{i}}^{\hat{l}} \d_{k}^{j} \,  \, \d^\a_{\; \, \b}
\, (N_1-N_2) \frac{\G^2(\frac12 - \e)}{16 \pi^{3-2\e}} \, \frac{1}{[(x-y)^2]^{1 - 2\e}}
\non \\
&~& \qquad= - \left( \frac{2\pi i}{k} \right) \, \d_I^J \d_{\hat{i}}^{\hat{l}} \d_{k}^{j} \,  \, \d^\a_{\; \, \b} \, (N_1-N_2) \frac{\G^2(\frac12 - \e) \G(\frac12 + \e)}{\G(1-2\e) 2^{3-2\e} \pi^{\frac32 -\e}} \, 
\int \frac{d^np}{(2\pi)^n}   \frac{e^{ip(x-y)} }{(p^2)^{\frac12 +\e}}
\eea 
and is proportional to the difference $(N_1-N_2)$ of the ranks of the gauge groups, hence it vanishes in the ABJM limit.

\section{Momentum space starting strings}\label{app:startingstrings}
In this appendix we list the starting expressions for the diagrams of Figure \ref{fig:4loops} in momentum space. These are  derived by applying the Feynman rules  of appendix \ref{app:FeynmanRules} using the Fourier transform \eqref{eq:Fourier}. Omitting a common factor $\left(\frac{2 \p }{k}\right)^4 \Tr (M_1M_2) \int \frac{d^{3-2\e}k_{1/2/3/4}}{(2\p)^{3-2\e}}$,  the full list of integrands, except for diagram (p) which is discussed in appendix \ref{app:vertex},  is given by 
\begin{align}
(a)  & = \frac{1}{k_4^2 (k_1-k_4)^2 (k_1-k_3)^2 (k_2-k_3)^2 }\non \\ &\bigg( \frac{8}{(i k_1\cdot v+\d)^3(i k_2\cdot v+\d)}+\frac{ \Tr(M_1M_2)}{ (i k_1\cdot v+\d)^2(i k_2\cdot v+\d)^2}\bigg) \\
(b) & = \frac{2}{k_4^2 (k_1-k_4)^2 (k_3-k_4)^2 (k_2-k_4)^2(i k_1\cdot v+\d)^2(i k_2\cdot v+\d)(i k_3\cdot v+\d)  }
\\
(c) & = -\frac{4}{k_4^2 (k_1-k_4)^2 (k_1-k_3)^2 (k_2-k_3)^2 (i k_1\cdot v+\d)^2(i k_2\cdot v+\d)^2} \\
(d) & =(c) -   \frac{8}{k_4^2 (k_1-k_4)^2 (k_1-k_3)^2 (k_2-k_3)^2 (i k_1\cdot v+\d)^3(i k_2\cdot v+\d)} \\
(e) & =\frac{ 2 i \, v_{\mu}((k_1-k_4)_{\nu}+(k_2-k_4)_{\nu})P^{\mu\nu}(k_1-k_2)}{k_4^2 (k_1-k_4)^2 (k_2-k_4)^2(i k_1\cdot v+\d)^2(i k_2\cdot v+\d)} \label{eq:e} \\
(f) & =  -  \frac{ 2 i \,v_{\mu}(k_{4 \nu}-(k_3-k_4)_{\nu})P^{\m\n}(k_3)}{k_4^2 (k_3-k_4)^2 (k_1-k_4)^2(i k_1\cdot v+\d)^2(i k_3\cdot v+\d) } \label{eq:f} \\
(g) & = - \frac{v^{\mu}v^{\nu}\varepsilon_{\mu\rho\eta}\varepsilon_{\rho\nu\l}(k_2-k_3)^{\eta}(k_1-k_2)^{\l}}{k_4^2 (k_1-k_2)^2 (k_2-k_3)^2 (k_3-k_4)^2 (k_1-k_4)^2}\non  \\ &\bigg(\frac{2}{(i k_1\cdot v+\d)^2(i k_2\cdot v+\d)(i k_3\cdot v+\d)}+\frac{1}{(i k_1\cdot v+\d)(i k_2\cdot v+\d)^2(i k_3\cdot v+\d)} \bigg)   \\ 
(h) & = - \frac{2 v^{\mu}v^{\nu}\varepsilon_{\mu\rho\eta}\varepsilon_{\rho\nu\l}k_3^{\eta}(k_2-k_3)^{\l}}{k_3^2 k_4^2  (k_2-k_3)^2 (k_2-k_4)^2 (k_1-k_4)^2(i k_1\cdot v+\d)^2(i k_2\cdot v+\d)(i k_3\cdot v+\d)}  \\ 
(i) & = \frac{v^{\mu}v^{\nu}\varepsilon_{\mu\rho\eta}\varepsilon_{\rho\nu\l}k_2^{\eta}k_3^{\l}}{k_3^2 k_2^2  (k_1-k_4)^2 (k_3-k_4)^2 (k_2-k_4)^2(i k_1\cdot v+\d)^2(i k_2 \cdot v+\d)(i k_3\cdot v+\d)}  \\  
(j) & = -\frac{ 4 i  \,v^{\mu}\varepsilon_{\rho\nu\l}\varepsilon_{\mu\nu\eta}(k_1-k_2)^{\eta}(k_3-k_4)^{\l}(k_3+k_4)^{\rho}}{k_3^2 k_4^2  (k_2-k_3)^2 (k_1-k_2)^2 (k_1-k_4)^2(k_3-k_4)^2(i k_1\cdot v+\d)^2(i k_2\cdot v+\d)}   \\ 
(k) & =  - \frac{4 i  \,v^{\mu}\varepsilon_{\rho\nu\l}\varepsilon_{\mu\nu\eta}k_3^{\eta}(k_2-k_4)^{\l}(k_1-k_2+k_1-k_4)^{\rho}}{k_3^2 k_4^2  (k_2-k_3)^2 (k_1-k_2)^2 (k_2-k_4)^2(k_1-k_4)^2(i k_1\cdot v+\d)^2(i k_3 \cdot v+\d)}   \\
(l) & =  \frac{ 2 i \,\varepsilon_{\l_1\l_2\l_3}\varepsilon_{\mu\l_1\eta}\varepsilon_{\l_2\nu\l_4}\varepsilon_{\l_3\rho\l_5} v^{\mu}(k_1-k_2)^{\eta}(k_1-k_3)^{\l_5}(k_2-k_3)^{\l_4}}{k_4^2 (k_1-k_2)^2  (k_1-k_3)^2(k_2-k_3)^2 (k_3-k_4)^2 (k_2-k_4)^2 (k_1-k_4)^2} \non  \\&\frac{\big[(k_3-k_4)^{\nu}(k_1-k_4)^{\rho}+(k_2-k_4)^{\nu}(k_3-k_4)^{\rho}+(k_3-k_4)^{\nu}(k_3-k_4)^{\rho}}{(i k_1\cdot v+\d)^2(i k_2 \cdot v+\d)} \non \\
& +(k_2-k_4)^{\nu}(k_1-k_4)^{\rho}\big] \\
(m) & = - \frac{2 i \,\varepsilon_{\l_1\l_2\l_3}\varepsilon_{\mu\l_1\eta}\varepsilon_{\l_2\nu\l_4}\varepsilon_{\l_3\rho\l_5} v^{\mu}k_3^{\eta}(k_2-k_3)^{\l_4}k_2^{\l_5}}{k_4^2 k_2^2 k_3^2 (k_1-k_4)^2  (k_2-k_3)^2(k_2-k_4)^2 (k_3-k_4)^2(i k_1\cdot v+\d)^2(i k_3 \cdot v+\d)} \non  \\&\big[-(k_2-k_4)^{\nu}k_4^{\rho}+(k_3-k_4)^{\nu}(k_2-k_4)^{\rho}+(k_2-k_4)^{\nu}(k_2-k_4)^{\rho}-(k_3-k_4)^{\nu}k_4^{\rho}\big] \\
(n) & = \frac{ 2 i\, \varepsilon_{\mu\nu\l_2}\varepsilon_{\rho\l_1\l_3}  v^{\mu}(k_1-k_2)^{\l_2}(k_3-k_4)^{\l_3}}{k_4^2 (k_1-k_3)^2 (k_1-k_2)^2 (k_2-k_3)^2 (k_3-k_4)^2 (k_2-k_4)^2 (k_1-k_4)^2} \non  \\&\frac{\big[
(k_1-k_3)^{\nu}(k_1-k_3)^{\rho}(k_2-k_4)^{\l_1}
+(k_2-k_3)^{\nu}(k_1-k_3)^{\rho}(k_2-k_4)^{\l_1}
}{(i k_1\cdot v+\d)^2(i k_2\cdot v+\d)} \non \\
& +(k_1-k_3)^{\nu}(k_1-k_4)^{\rho}(k_2-k_4)^{\l_1}
+(k_1-k_3)^{\nu}(k_1-k_3)^{\rho}(k_2-k_3)^{\l_1}
 \non \\
&+(k_2-k_3)^{\nu}(k_1-k_3)^{\rho}(k_2-k_3)^{\l_1} +(k_1-k_3)^{\nu}(k_1-k_4)^{\rho}(k_2-k_3)^{\l_1}
 \non \\& +(k_2-k_3)^{\nu}(k_1-k_4)^{\rho}(k_2-k_3)^{\l_1} +(k_2-k_3)^{\nu}(k_1-k_4)^{\rho}(k_2-k_4)^{\l_1} \big]\\
(o) & =  \frac{2 i\,  \varepsilon_{\mu\nu\l_2}\varepsilon_{\rho\l_1\l_3}  v^{\mu}k_3^{\l_2}(k_2-k_4)^{\l_3}}{k_4^2 k_2^2 k_3^2 (k_1-k_4)^2 (k_2-k_3)^2 (k_2-k_4)^2 (k_3-k_4)^2 (i k_1\cdot v+\d)^2(i k_3\cdot v+\d)}  \\&\big[
-k_2^{\nu}k_2^{\rho}(k_3-k_4)^{\l_1}
-(k_2-k_3)^{\nu} k_2^{\rho}(k_3-k_4)^{\l_1}
 -k_2^{\nu}k_4^{\rho}(k_3-k_4)^{\l_1}\non \\
&-(k_2-k_3)^{\nu}k_4^{\rho}(k_3-k_4)^{\l_1} +k_2^{\nu}k_2^{\rho}(k_2-k_3)^{\l_1} +(k_2-k_3)^{\nu}k_2^{\rho}(k_2-k_3)^{\l_1} \non\\
  &+ k_2^{\nu}k_4^{\rho}(k_2-k_3)^{\l_1} +(k_2-k_3)^{\nu}k_4^{\rho}(k_2-k_3)^{\l_1} \big] \non 
\end{align}
where $P^{\mu\nu}(k)$ in equations \eqref{eq:e} and   \eqref{eq:f} stands for the 1 loop gauge propagator with momentum $k$, which can be read by \eqref{1vector}.

\section{Master integrals definitions and expansions}\label{app:masters}

We define the (Euclidean) HQET planar integrals at two and four loops by the following products of propagators ($d=3-2\epsilon$)
\begin{align}\label{eq:masterintegrals}
\text{two loops:} & \,\,\, G_{a_1,a_2, a_3, a_4} \equiv \int \frac{d^dk_1\, d^dk_2}{(2\pi)^{2d}}\, \frac{1}{P_1^{a_1}\, P_2^{a_2}\, P_5^{a_3} P_6^{a_4}\, P_9^{a_5}} \nonumber\\
\text{four loops:} &  \,\,\,G_{a_1,\dots,a_{14}} \equiv \int \frac{d^dk_1\, d^dk_2\, d^dk_3\, d^dk_4}{(2\pi)^{4d}}\, \prod_{i=1}^{14}\frac{1}{P_i^{a_i}}
\end{align}
where the explicit propagators read
\begin{align}
& P_1 = (2k_1\cdot \tilde v+1) ,\quad P_2 = (2k_2\cdot \tilde v+1) ,\quad P_3 = (2k_3\cdot \tilde v+1) ,\quad P_4 = (2k_4\cdot \tilde v+1) \nonumber\\&
P_5 = k_1^2 ,\qquad P_6 = k_2^2 ,\qquad P_7 = k_3^2 ,\qquad P_8 = k_4^2 \nonumber\\&
P_9 = (k_1-k_2)^2 ,\qquad P_{10} = (k_1-k_3)^2 ,\qquad P_{11} = (k_1-k_4)^2  \nonumber\\&
P_{12} = (k_2-k_3)^2 ,\qquad P_{13} = (k_2-k_4)^2 ,\qquad P_{14} = (k_3-k_4)^2
\end{align}
and $\tilde v^2=-1$.

The cusp computation presented in this note requires the expansion and evaluation of the 21 master integrals of Figure \ref{fig:master} to certain orders in $\epsilon$, depending on the integral.
Here we provide the relevant expansions needed for the calculation.

A subset of the master integrals can be evaluated exactly in terms of lower order ones.
In particular we can use the following bubble integrals
\begin{align}
\begin{minipage}{3cm}\vspace{-0.45cm}
\includegraphics[scale=0.35]{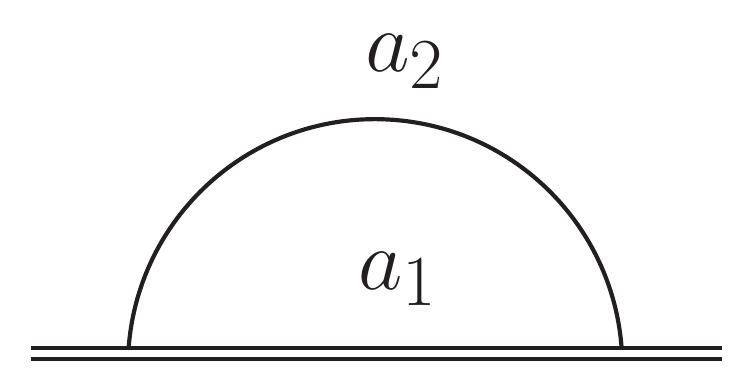}
\end{minipage} &= I(a_1,a_2) = G_{a_1,a_2}  = \frac{1}{(4 \pi)^{d/2}}\frac{\Gamma(a_1+ 2 a_2-d) \Gamma(d/2-a_2)}{\Gamma(a_1)\Gamma(a_2)}
\\[4mm]
\begin{minipage}{3cm}\vspace{-0.45cm}
\includegraphics[scale=0.525]{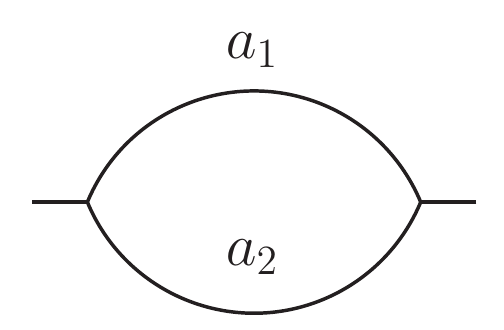}
\end{minipage} &= B_{a_1,a_2}  = \frac{1}{(4 \pi)^{d/2}}\frac{\Gamma(d/2-a_1)\Gamma(d/2-a_2)\Gamma(a_1 + a_2 - d/2)}{\Gamma(a_1)\Gamma(a_2)\Gamma(d - a_1 - a_2 )}
\end{align}
and obtain for instance (we drop the $4\pi$ normalization and $\gamma_E$ factors in what follows)
\begin{equation}
\raisebox{-4mm}{\includegraphics[scale=0.4]{MI5}} = I(1,1)\, B^2(1,1)\, I(2\epsilon,1+2\epsilon) = -\frac{\pi ^4}{8 \epsilon }-\frac{1}{2} \pi ^4 (2+3\log 2)+{\cal O}(\epsilon)
\end{equation}
Another subset of master integrals is obtained by lower order topologies supplemented by an additional external propagator, such as $G_{1, 1, 1, 0, 0, 0, 0, 1, 0, 1, 1, 1, 1, 0}$.
These integrals all factorize and can be evaluated using the master integrals of \cite{Bianchi:2017svd}, for instance
\begin{equation}
\raisebox{-4mm}{\includegraphics[scale=0.4]{MI15}} = I(2+6\epsilon,1)\, \raisebox{-3mm}{\includegraphics[scale=0.4]{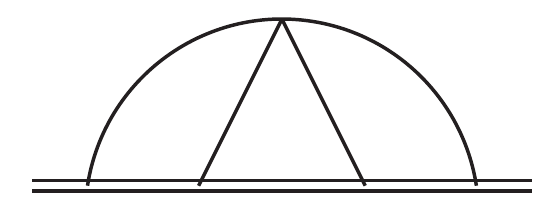}} = \frac{2 \pi ^4}{3}+{\cal O}(\epsilon)
\end{equation}
Other integrals can be mapped to two- and three-loop topologies with non-integer indices, after integrating bubble subtopologies, such as
\begin{equation}
\raisebox{-4mm}{\includegraphics[scale=0.4]{MI12}} = B(1,1)\, I(1,1)\, \raisebox{-6.5mm}{\includegraphics[scale=0.4]{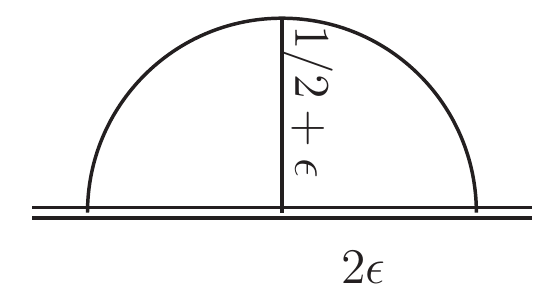}} 
\end{equation}
These cases can be dealt with by deriving expressions for general indices using the Gegenbauer polynomial technique (GPXT) \cite{Chetyrkin:1980pr} or a Mellin-Barnes (MB) representation.

\noindent
The only genuinely four-loop integrals are $G_{0, 1, 1, 0, 0, 0, 1, 1, 1, 1, 1, 0, 1, 0}$ and $G_{1, 0, 1, 0, 0, 1, 0, 1, 0, 1, 1, 1, 1, 0}$ (see eqs \ref{4Lmast1} and \ref{4Lmast2}).
For these we derived a 6-fold MB representation. After $\epsilon$ expansion to the required order and some MB integral gymnastics we were able to reduce all the relevant expressions to one-fold integrals which could be evaluated using the Barnes lemmas and their corollaries, or evaluated directly, such as
\begin{equation}
\int_{-i\infty}^{+i\infty} \frac{du}{2\pi i}\, \frac{\Gamma(-u)\Gamma(1/2-u)\Gamma^3(1/2+u)}{\Gamma(1+u)} = {}_3F_2\left( \begin{array}{ccccc} \frac12 & & \frac12 & & \frac12 \\ & \frac32 & & 1 & \end{array}, 1 \right) = 8\, C
\end{equation}
where $C$ is the Catalan constant, providing full analytic results.
Altogether, the expansions of the master integrals read
\begin{align}
\raisebox{-4mm}{\includegraphics[scale=0.4]{MI1}} &= -\frac{\pi ^2}{48 \epsilon }-\frac{1}{36} \pi ^2 (11+6 \log 2) \nonumber\\& -\frac{1}{432} \left(\pi ^2 \left(57 \pi ^2+8 (170+6 \log 2 (22+6 \log 2))\right)\right) \epsilon +{\cal O}\left(\epsilon ^2\right) \\
\raisebox{-1.5mm}{\includegraphics[scale=0.4]{MI2}} &= \frac{\pi ^2}{16 \epsilon ^2}+\frac{\pi ^2 (2+2 \log 2)}{4 \epsilon }\nonumber\\& +\frac{1}{48} \pi ^2 \left(11 \pi ^2+24 (6+2 \log 2 (4+2 \log 2))\right)+{\cal O}\left(\epsilon ^1\right) \\
\raisebox{-1.5mm}{\includegraphics[scale=0.4]{MI4}} &= -\frac{\pi ^2}{12 \epsilon ^3}-\frac{\pi ^2 (3+4 \log 2)}{6 \epsilon ^2} \nonumber\\& -\frac{\pi ^2 \left(11 \pi ^2+12 (9+4 \log 2 (3+2 \log 2))\right)}{36 \epsilon }
+{\cal O}\left(\epsilon ^0\right) \\
\raisebox{-4mm}{\includegraphics[scale=0.4]{MI21}} &= -\frac{3 \pi ^2}{32 \epsilon ^3}-\frac{3 \left(\pi ^2 (1+\log 2)\right)}{4 \epsilon ^2} \nonumber\\& -\frac{\pi ^2 \left(13 \pi ^2+96 \left(2+2 \log 2+\log ^2 2\right)\right)}{32 \epsilon }
+{\cal O}\left(\epsilon ^0\right) \\
\raisebox{-4mm}{\includegraphics[scale=0.4]{MI6}} &= \frac{3 \pi ^2}{64 \epsilon ^2}+\frac{3 \pi ^2 (5+4 \log 2)}{32 \epsilon } \nonumber\\& +\frac{3}{64} \pi ^2 \left(5 \pi ^2+8 (10+2 \log 2 (5+2 \log 2))\right)+{\cal O}\left(\epsilon ^1\right) \\
\raisebox{-4mm}{\includegraphics[scale=0.4]{MI5}} &= -\frac{\pi ^4}{8 \epsilon }-\frac{1}{2} \pi ^4 (2+3 \log 2)+{\cal O}\left(\epsilon ^1\right) \\
\raisebox{-4mm}{\includegraphics[scale=0.4]{MI7}} &= \frac{\pi ^2}{16 \epsilon ^2}+\frac{\pi ^2 (3+2 \log 2)}{4 \epsilon } \nonumber\\& +\frac{1}{48} \pi ^2 \left(17 \pi ^2+24 (14+2 \log 2 (6+2 \log 2))\right)+{\cal O}\left(\epsilon ^1\right) \\
\raisebox{-4mm}{\includegraphics[scale=0.4]{MI8}} &= \frac{\pi ^2}{24 \epsilon ^2}+\frac{\pi ^2 (9+8 \log 2)}{24 \epsilon } \nonumber\\& +\frac{1}{72} \pi ^2 \left(189+13 \pi ^2+216 \log 2+96 \log ^2 2\right)+{\cal O}\left(\epsilon ^1\right) \\
\raisebox{-4mm}{\includegraphics[scale=0.4]{MI9}} &= -\frac{\pi ^2}{8 \epsilon ^2}-\frac{\pi ^2 (7+4 \log 2)}{4 \epsilon } \nonumber\\& -\frac{1}{24} \pi ^2 \left(11 \pi ^2+12 (37+4 \log 2 (7+2 \log 2))\right)+{\cal O}\left(\epsilon ^1\right) \\
\raisebox{-4mm}{\includegraphics[scale=0.4]{MI20}} &= -\frac{\pi ^4}{2 \epsilon }-3 \left(\pi ^4 (-1+2 \log 2)\right)+{\cal O}\left(\epsilon ^1\right) \\
\raisebox{-4mm}{\includegraphics[scale=0.4]{MI11}} &= \frac{\pi ^4}{48 \epsilon ^2}+\frac{\frac{1}{2} \pi ^4 \log 2-\frac{7 \pi ^2 \zeta (3)}{8}}{\epsilon }+{\cal O}\left(\epsilon ^0\right) \\
\raisebox{-4mm}{\includegraphics[scale=0.4]{MI15}} &= \frac{2 \pi ^4}{3}+{\cal O}\left(\epsilon ^1\right) \\
\raisebox{-4mm}{\includegraphics[scale=0.4]{MI3}} &= \frac{\pi ^4}{16 \epsilon ^2}+\frac{\pi ^4 \log 2-\frac{7 \pi ^2 \zeta (3)}{8}}{\epsilon }+{\cal O}\left(\epsilon ^0\right) \\
\raisebox{-4mm}{\includegraphics[scale=0.4]{MI12}} &= \frac{\pi ^2}{24 \epsilon ^3}+\frac{\pi ^2 (5+4 \log 2)}{12 \epsilon ^2} \nonumber\\& +\frac{\pi ^2 \left(13 \pi ^2+12 (25+4 \log 2 (5+2 \log 2))\right)}{72 \epsilon }+{\cal O}\left(\epsilon ^0\right) \\
\raisebox{-4mm}{\includegraphics[scale=0.4]{MI13}} &= {\cal O}\left(\frac{1}{\epsilon }\right) \\
\raisebox{-4mm}{\includegraphics[scale=0.4]{MI10}} &= -\frac{\pi ^4}{2 \epsilon }+{\cal O}\left(\epsilon ^0\right) \\
\raisebox{-4mm}{\includegraphics[scale=0.4]{MI16}} &= \frac{2 \pi ^4}{3 \epsilon }+{\cal O}\left(\epsilon ^0\right) \\
\raisebox{-4mm}{\includegraphics[scale=0.4]{MI17}} &= {\cal O}\left(\frac{1}{\epsilon }\right) \\
\raisebox{-4mm}{\includegraphics[scale=0.4]{MI18}} &= \frac{2 \pi ^4}{3}+{\cal O}\left(\epsilon ^1\right) \\
\raisebox{-4mm}{\includegraphics[scale=0.4]{MI19}} &= \frac{\pi ^2}{2 \epsilon ^2}+\frac{4 \pi ^2 \log 2}{\epsilon }+\frac{1}{6} \pi ^2 \left(-84+7 \pi ^2+96 \log ^2 2\right)+{\cal O}\left(\epsilon ^1\right) \label{4Lmast1}\\
\raisebox{-4mm}{\includegraphics[scale=0.4]{MI14}} &= \frac{\pi ^2}{2 \epsilon ^2}+\frac{4 \pi ^2 \log 2}{\epsilon }+\frac{1}{6} \pi ^2 \left(-180+23 \pi ^2+96 \log ^2 2\right)+{\cal O}\left(\epsilon ^1\right)  \label{4Lmast2}
\end{align}
where an overall factor $e^{-4 \gamma_E \epsilon}/(4\pi)^{2d}$ is omitted.

\section{Results for the four-loop diagrams}\label{app:diagrams}

Here we list the results for the diagrams of Figure \ref{fig:4loops}.
A common factor $\left(\frac{e^{-4\epsilon\gamma_E}}{k(4\pi)^{d/2}}\right)^4$ is understood.

\begin{align}
(a) &= \frac{8 \pi ^2 C_{\theta }^2 \left(C_{\theta }^2-2\right) N_1^2 N_2^2}{\epsilon ^2}+\frac{32 \pi ^2 C_{\theta }^2 \left((-3+6 \log 2) C_{\theta }^2-12 \log 2\right) N_1^2 N_2^2}{3 \epsilon }+{\cal O}\left(\epsilon ^0\right)
\\
(b) &= \frac{4 \pi ^2 C_{\theta }^2 N_1^3 N_2}{\epsilon ^3}+\frac{32 \pi ^2 \log 2 C_{\theta }^2 N_1^3 N_2}{\epsilon ^2}+\frac{4 \pi ^2 \left(13 \pi ^2+96 \log^2 2\right) C_{\theta }^2 N_1^3 N_2}{3 \epsilon }+{\cal O}\left(\epsilon ^0\right)
\\
(c) &= -\frac{16 \pi ^2 C_{\theta }^2 N_1^2 N_2^2}{\epsilon ^2}-\frac{16 \pi ^2 (-7+8 \log 2) C_{\theta }^2 N_1^2 N_2^2}{\epsilon }+{\cal O}\left(\epsilon ^0\right)
\\
(d) &= \frac{16 \pi ^2 C_{\theta }^2 N_1^2 N_2^2}{\epsilon ^2}+\frac{16 \pi ^2 (1+8 \log 2) C_{\theta }^2 N_1^2 N_2^2}{\epsilon }+{\cal O}\left(\epsilon ^0\right)
\\
(e) &= \frac{128 \pi ^2 C_{\theta }^2 N_1^2 N_2^2}{\epsilon }+{\cal O}\left(\epsilon ^0\right)
\\
(f) &= -\frac{32 \pi ^2 \left(-4+\pi ^2\right) C_{\theta }^2 N_1^2 N_2^2}{\epsilon }+{\cal O}\left(\epsilon ^0\right)
\\
(g) &= -\frac{2 \pi ^2 C_{\theta }^2 N_1^3 N_2}{\epsilon ^3}-\frac{2 \pi ^2 (-1+8 \log 2) C_{\theta }^2 N_1^3 N_2}{\epsilon ^2}\nonumber\\&+\frac{4 \pi ^2 \left(9-7 \pi ^2+12 \log 2-48 \log^2 2\right) C_{\theta }^2 N_1^3 N_2}{3 \epsilon }+{\cal O}\left(\epsilon ^0\right)
\\
(h) &= -\frac{2 \pi ^2 C_{\theta }^2 N_1^3 N_2}{\epsilon ^3}-\frac{2 \pi ^2 (-1+8 \log 2) C_{\theta }^2 N_1^3 N_2}{\epsilon ^2}\nonumber\\&+\frac{2 \pi ^2 \left(18-17 \pi ^2+24 (1-4 \log 2) \log 2\right) C_{\theta }^2 N_1^3 N_2}{3 \epsilon }+{\cal O}\left(\epsilon ^0\right)
\\
(i) &= -\frac{2 \pi ^4 C_{\theta }^2 N_1^3 N_2}{3 \epsilon }+{\cal O}\left(\epsilon ^0\right)
\\
(j) &= \frac{8 \pi ^4 C_{\theta }^2 N_1^2 N_2^2}{3 \epsilon }+{\cal O}\left(\epsilon ^0\right)
\\
(k) &= -\frac{8 \pi ^4 C_{\theta }^2 N_1^2 N_2^2}{3 \epsilon }+{\cal O}\left(\epsilon ^0\right)
\\
(l) &= -\frac{4 \pi ^4 C_{\theta }^2 N_1^3 N_2}{3 \epsilon }+{\cal O}\left(\epsilon ^0\right)
\\
(m) &= \frac{4 \pi ^4 C_{\theta }^2 N_1^3 N_2}{3 \epsilon }+{\cal O}\left(\epsilon ^0\right)
\\
(n) &= -\frac{8 \pi ^4 C_{\theta }^2 N_1^2 N_2^2}{3 \epsilon }+{\cal O}\left(\epsilon ^0\right)
\\
(o) &= \frac{8 \pi ^4 C_{\theta }^2 N_1^2 N_2^2}{3 \epsilon }+{\cal O}\left(\epsilon ^0\right)
\\
(p) &= -\frac{4 \pi ^2 C_{\theta }^2 N_1^2 N_2 \left(N_1+4 N_2\right)}{\epsilon ^2}+\frac{4 \pi ^2 C_{\theta }^2 N_1 N_2}{3 \epsilon } \left(3 \left(\pi ^2-8 \log 2-6\right) N_1^2 \right.\nonumber\\&\left.+4 \left(\pi ^2-6 (7+4 \log 2)\right) N_2 N_1-4 \pi ^2 N_2^2\right)+{\cal O}\left(\epsilon ^0\right) 
\end{align}

\section{Scalar bubble corrections}\label{app:vertex}

\begin{figure}[h]
\centering
\includegraphics[scale=0.3]{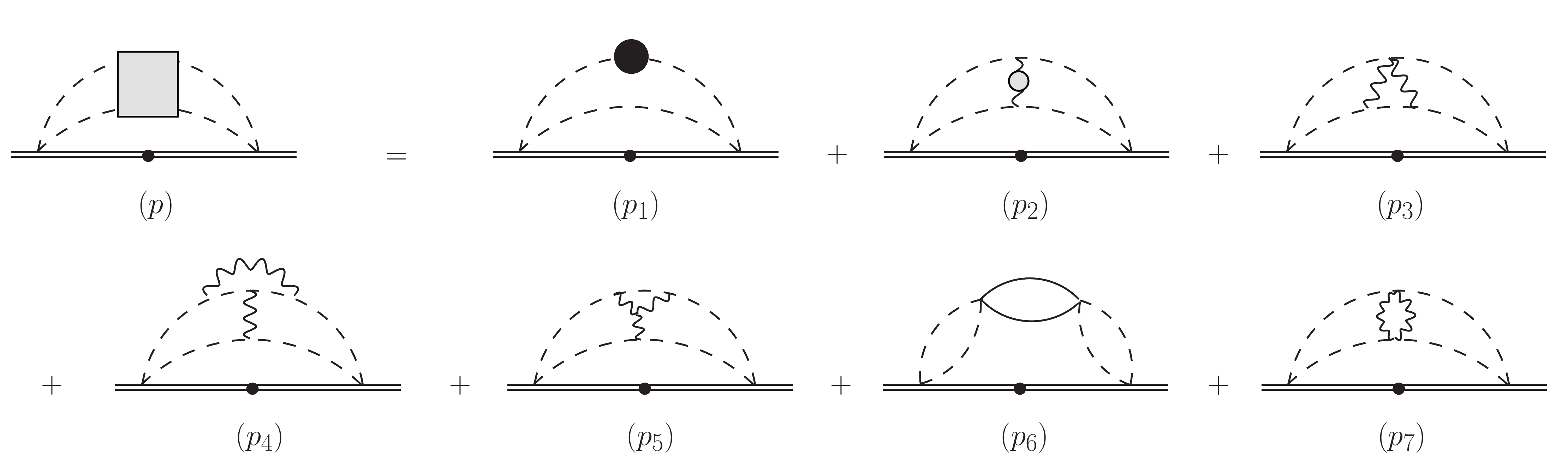}
\caption{Scalar bubble corrections} \label{diag_p}
\end{figure}

Diagram (p) of Figure \ref{fig:4loops} represents collectively the internal corrections to the scalar bubble. The non-vanishing  contributions are listed in Figure  \ref{diag_p}.

\noindent
To compute diagram $(p_1)$ we also need the expression for the 2-loop correction to the scalar propagator, which was given for instance in \cite{Minahan:2009wg}. 
Altogether, the various contributions from diagram (p) to the cusp expectation value read
\begin{align}
(p_1) &= -\frac{4 \pi ^2 N_1 N_2 \left(N_1^2+4 N_2 N_1+N_2^2\right) C_{\theta }^2}{\epsilon ^2}
\\&
+\frac{4 \pi ^2 N_1 N_2 C_{\theta }^2 \left((N_1^2+N_2^2) \left(\pi ^2-8 \log 2-6\right)+4 N_2 N_1 \left(\pi ^2-8 \log 2-22\right)\right)}{\epsilon } +{\cal O}\left(\epsilon ^0\right)\nonumber
\\
(p_2) &= -\frac{16 \pi ^2 \left(\pi ^2-12\right) N_1^2 N_2^2 C_{\theta }^2}{\epsilon }+{\cal O}\left(\epsilon ^0\right)
\\
(p_3) &= -\frac{4 \pi ^2 \left(\pi ^2-12\right) N_1 N_2^3 C_{\theta }^2}{\epsilon }+{\cal O}\left(\epsilon ^0\right)
\\
(p_4) &= \frac{16 \pi ^2 \left(\pi ^2-12\right) N_1^2 N_2^2 C_{\theta }^2}{3 \epsilon }+{\cal O}\left(\epsilon ^0\right)
\\
(p_5) &= \frac{8 \pi ^2 \left(\pi ^2-12\right) N_1 N_2^3 C_{\theta }^2}{3 \epsilon }+{\cal O}\left(\epsilon ^0\right)
\\
(p_6) &= -\frac{8 \pi ^4 N_1 N_2^3 C_{\theta }^2}{\epsilon }+{\cal O}\left(\epsilon ^0\right)
\\
(p_7) &= \frac{4 \pi ^2 N_1 N_2^3 C_{\theta }^2}{\epsilon ^2}+\frac{8 \pi ^2 N_1 N_2^3 (1+4 \log 2) C_{\theta }^2}{\epsilon }+{\cal O}\left(\epsilon ^0\right)
\end{align}

\bibliographystyle{JHEP}

\bibliography{biblio}

\end{document}